\documentclass[twocolumn]{aastex631}
\usepackage{amsmath}
\usepackage{xcolor, colortbl}
\definecolor{vds_color}{HTML}{9A9A9A} 
\definecolor{ds_color}{HTML}{BCBCBC}
\definecolor{s_color}{HTML}{DDDDDD}
\definecolor{h_color}{HTML}{dddde3}
\definecolor{darkgreen}{HTML}{2dbb2d}

\begin{document}
\definecolor{pcolor}{HTML}{E7D1FF}  
\title{Physical Properties of Molecular Clouds in
the Overlap Region of the Merging Antennae Galaxies}
\author[0000-0002-0932-9879]{Grace Krahm}
\affiliation{Department of Physics and Astronomy, Agnes Scott College, Decatur, GA 30030, USA}
\affiliation{National Radio Astronomy Observatory, 520 Edgemont Road, Charlottesville, VA 22903, USA}

\author[0000-0001-9338-2594]{Molly K. Finn}
\affiliation{Department of Astronomy, University of Virginia, Charlottesville, VA 22904, USA}

\author[0000-0002-4663-6827]{Remy Indebetouw}
\affiliation{Department of Astronomy, University of Virginia, Charlottesville, VA 22904, USA}
\affiliation{National Radio Astronomy Observatory, 520 Edgemont Road, Charlottesville, VA 22903, USA}

\author[0000-0001-8348-2671]{Kelsey E. Johnson}
\affiliation{Department of Astronomy, University of Virginia, Charlottesville, VA 22904, USA}

\author[0000-0001-7877-7942]{Julia Kamenetzky}
\affiliation{Department of Physics, Westminster College, 1840 S 1300 E. Salt Lake City, UT 84105 USA}

\author[0000-0003-0618-8473]{Ashley Bemis}
\affiliation{Leiden University, Rapenburg 70, 2311 EZ, Leiden, The Netherlands}

\begin{abstract}
As the closest major galaxy merger and home to thousands of super star clusters (SSCs), the
Antennae Galaxies (NGC 4038 and NGC 4039) are an important location to study the molecular
clouds at sites of vigorous star formation. We cataloged giant molecular clouds (GMCs) in the region where the two galaxies overlap using high-resolution ($\sim 0.1" \sim 10$pc) Atacama Large
Millimeter/submillimeter Array (ALMA) observations of the $\mathrm{^{12}CO(2-1) \:and\: ^{13}CO(2-1)}$
emission lines. Of the 72 individual GMCs identified in the overlap region, 17 are within uncertainties of having the necessary mass, pressure, and size needed to form super star clusters (SSCs). Of those 17 GMCs, only one has significant ionizing radiation, indicating that the birth environments are likely still intact in the 16 other GMCs. We compared the physical properties calculated from $\mathrm{^{12}CO(2-1)}$ GMC data with observations of 10 other galaxies obtained using the same emission line and similar resolution. Compared to other sources in this sample, the GMCs from the Antennae, as well as in other starbursts and in the centers of galaxies, have the highest luminosities, surface densities, and turbulent pressures. The GMCs in starbursts and at the centers of galaxies also have large linewidths, although the linewidths in the Antennae are \textcolor{black}{among the widest}. These comparative results also indicate that the Antennae GMCs have the highest virial parameters despite their high
densities.
\end{abstract}

\section{Introduction}\label{sec:intro}

Due to their old ages and abundance throughout the observed universe, 
globular clusters have been used as probes to study stellar and galactic formation throughout cosmic time \citep{1976ApJ...209..418H,2010ARA&A..48..431P,2010ApJ...718.1266M,2015MNRAS.454.1658K,2019ApJ...874..120F}. In order to better understand how globular clusters form, we must look at young star clusters and the molecular gas surrounding them.

Super star clusters (SSCs) are young massive star clusters that are extremely dense, with average ranges of radii and masses being 1-5 pc \citep{2010RSPTA.368..867L, 2012ApJ...750..136W} and $\mathrm{10^{4}-10^{6} M_{\odot}}$ \citep{2012ApJ...750..136W} respectively. Due to the fact that SSCs have sizes and masses similar to globular clusters, it is believed that many SSCs are early-stage globular clusters \citep{1995AJ....110.2665M,2007ApJ...663..844M}. Consequently, the prevailing hypothesis is that the processes that formed globular clusters are still occurring in some galaxies, although only in extreme environments, rather than being confined to the early universe \citep{1994ApJ...433...65O,2001AJ....122.1888A,2008ApJ...679.1272M}.

Early work studying SSCs was primarily carried out using optical and UV wavelengths, and was limited to studying SSCs older than a few million years \citep{1993AJ....106.1354W,1994ApJ...433...65O,1995AJ....109..960W,1999AJ....118.1551W,2000AJ....119.2146J,2000AJ....120.1273J}. In order to study earlier stages of SSC evolution when the clusters are still embedded in their birth material, there was a push in the late 1990s to probe the birth environments of SSCs using longer wavelengths (e.g. \citet{1998AJ....116.1212T}). \citet{1999ApJ...527..154K} termed the compact thermal radio sources associated with natal SSCs ``ultradense HII regions" (UDHIIs) due to their similar properties to Galactic ultracompact HII regions (UCHIIs). Since then, UDHIIs have been found in numerous other starbursts \citep{2000AJ....120..244B,2000ApJ...532L.109T,2000A&A...358...95T,2001ApJ...559..864J,2003AJ....126..101J,2004AJ....128.1552B,2008AJ....135.2222R,2009A&A...504..415S,2013MNRAS.435..400U,2014AJ....147...43K}.

As instrumentation became available, work to characterize natal SSCs continued in the infrared \citep{2002AJ....123..772V,2004AJ....128..610J,2006ApJ...653.1129B,2006A&A...446..877M,2007AJ....134.1522J,2007ApJ...671..333K,2013ApJS..209...26F}. When ALMA came online, sensitive, high-resolution imaging of the natal molecular clouds was finally possible, enabling numerous studies \citep{2012A&A...538L...9H,2013ApJ...774...73I,2015ApJ...801...25L,2015ApJ...806...35J,2017ApJ...846...73T,2018ApJ...853..125J,Miura_2018,2019ApJ...877..135N,2019ApJ...874..120F,2020ApJ...903...50E,2020ApJ...899...94R, 10.1093/mnras/stab085,2022arXiv220611242F,2022ApJ...928...57H}.

Multiwavelength observations and numerical models have allowed several constraints to be placed on the physical properties of SSC birth environments. In order to form an SSC, and by extension a globular cluster, the progenitor giant molecular cloud (GMC) would need a radius of $< 25$ pc and a mass \textcolor{black}{$>10^{6}\:M_{\odot}$} \citep{2015ApJ...806...35J}, and an external pressure of $\mathrm{P_{e}/k>10^{7}}$ K $\mathrm{cm^{-3}}$ \citep{1997ApJ...480..235E}. This is calculated by assuming the typical globular cluster has a half-light radius of $\mathrm{<}$ 10 pc \citep{1991ApJ...375..594V}, a stellar mass of $10^{5} M_{\odot}$ \citep{1994ApJ...429..177H}, and a star formation efficiency (SFE) of $20-50 \%$ \citep{2001AJ....122.1888A}. These parameters provide guidance on selection criteria for identifying molecular clouds with the potential to form SSCs.  

At a distance of 22 Mpc \citep{2008AJ....136.1482S}, the merging Antennae galaxies (NGC 4038/39) are of particular interest as the closest major galaxy merger. Their interaction has resulted in a dense region under high pressure where the galaxies overlap in their merger referred to as the overlap region. \citep{2021PASJ...73S..35T}. This overlap region is ideal for the formation of SSCs, as evidenced by the discovery of thousands of SSCs \citep{2007ApJ...668..168G,2010AJ....140...75W} and a pre-SSC cloud called the Firecracker \citep{2014ApJ...795..156W,2015ApJ...806...35J,2019ApJ...874..120F} in the region. By calculating the physical properties of molecular clouds in the Antennae overlap region and comparing them to \textcolor{black}{GMCs in} both SSC-forming and non-SSC-forming galaxies alike, we can determine the relative importance of different properties in the SSC formation process.  

In this paper, we present high-resolution ALMA observations of $\mathrm{^{12}CO(2-1)}$, $\mathrm{^{12}CO(3-2)}$, and $\mathrm{^{13}CO(2-1)}$ to characterize the physical conditions of GMCs in the Antennae overlap region and determine how they compare to GMCs in other galaxies (both SSC-forming and non-SSC-forming). In Section \ref{sec:obs} we discuss the observations of the Antennae overlap used in this analysis. Section \ref{sec:structure} describes the different outlier detection methods we used to identify individual GMCs and characterize the overall structure of the overlap region, which we then use to calculate the physical properties of GMCs in the overlap in Section \ref{sec:props}. Section \ref{sec:antprops} discusses the distributions and average values of select physical properties as well as the SSC-forming capabilities of clumps that significantly deviate from the normal distribution. In Section \ref{sec:companalasys} we compare the physical properties of GMCs in the Antennae to those in other starbursting and non-starbursting environments. The implications of these comparisons are discussed in Section \ref{sec:discuss}, and Section \ref{sec:conc} summarizes our findings in this work.

\section{Observations}\label{sec:obs}
We used ALMA data from projects 2015.1.00977.S and 2016.1.00924.S during Cycles 3 and 4. This includes Band 6 and 7 observations of the $\mathrm{^{12}CO(2-1)}$, $\mathrm{^{13}CO(2-1)}$, and $\mathrm{^{12}CO(3-2)}$ emission lines focused on the Antennae overlap region which are summarized in Table \ref{tab:obslog}. 

\begin{figure}
\centering
\includegraphics[width=.47\textwidth]{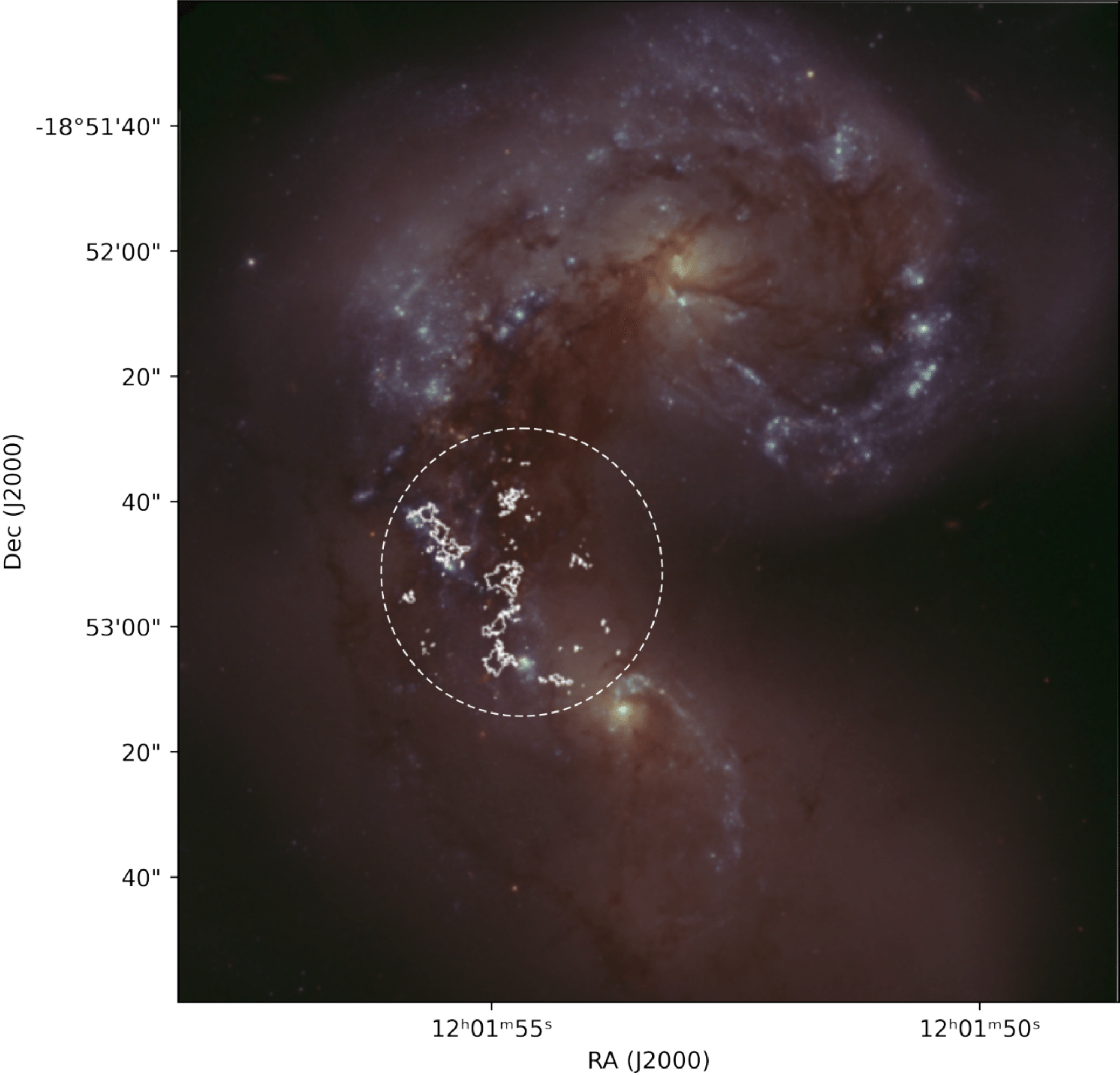}
\caption{Three-color NASA/ESA Hubble Space Telescope image of the Antennae galaxies where red is F814W, green is F550M, and blue is F435W. Plotted overtop are the $\mathrm{^{12}CO(2-1)}$ moment 0 contours ($\mathrm{3\sigma}$, white) of the Antennae overlap used in this work. The dashed white circle denotes the ALMA Band 6 primary beam (FWHM 25").}
\label{fig:antobs}
\end{figure}

We used the J1256-0547 and J1037-2934 flux calibrators for flux calibration and estimated a flux uncertainty of 10\% due to the source variability. We also used J1256-0547, J1229+0203, and J1037-2934 to calibrate the bandpass and J1215-1731 for phase calibration. All calibration and reduction of the observations were done using the CASA 4.7.2 pipeline. A more in-depth description of the observation parameters and calibration is available in \citet{2019ApJ...874..120F}.

In order to use multiple emission lines in the analysis, we convolved the data cubes to the same resolution as the $\mathrm{^{13}CO(2-1)}$ data cube, which had the lowest resolution. A primary beam correction was also performed using CASA. 

In addition to the CO emission lines from ALMA, we also used archival 3.6 cm (8.46 GHz) radio continuum data (project AN074)taken \text{in 1997} with the Karl G. Jansky Very Large Array (VLA) \textcolor{black}{at} X-band in the A configuration \citep{2014ApJ...795..156W,2015ApJ...806...35J}. The noise and resolution are shown in Table \ref{tab:obslog}.

\begin{deluxetable}{cccc}
\tabletypesize{\scriptsize}
\tablecaption{Observations \& Data Cube Parameters\label{tab:obslog}}
\tablehead{\colhead{Line} & \colhead{Beam} & \colhead{$\mathrm{\sigma_{rms}}$} & \colhead{Channel Width} \\ 
\colhead{} & \colhead{(\arcsec x \arcsec)} & \colhead{(K)} & \colhead{($\mathrm{km\:s^{-1}}$)} } 
\startdata
$\mathrm{^{12}CO(2-1)}$ & 0.09 x 0.12 &  1.74 & 5 \\
$\mathrm{^{12}CO(3-2)}$ & 0.15 x 0.16 &  0.48 & 5 \\
$\mathrm{^{13}CO(2-1)}$ & 0.17 x 0.18 & 1.29 & 5 \\
3.6 cm continuum&0.42 x 0.65&2.32&...
\enddata
\tablecomments{Both the $\mathrm{^{12}CO(2-1)}$ and $\mathrm{^{12}CO(3-2)}$ lines were later convolved to the same 0.166" x 0.180" resolution as $\mathrm{^{13}CO(2-1)}$ for analysis.}
\end{deluxetable}

\section{Structure Decomposition}\label{sec:structure}
We used two different methods to decompose the structure of the Antennae overlap region, which we refer to throughout as clumps and dendrogram structures. Clump-finding algorithms separate the emission by identifying local maxima and the surrounding pixels above a minimum intensity threshold and assigning them into singular clumps that do not overlap with one another.

Clump-finding algorithms also have some limitations in representing the morphology of the emission since they cannot capture the hierarchical nature of molecular clouds and are biased toward finding circular, beam-sized clouds. Dendrogram algorithms eliminate these issues by splitting the emission into a hierarchy of structures. While dendrogram structures better characterize the size scales and hierarchical nature of GMCs, they can only be used for non-counting-based analysis since areas of emission are counted multiple times. Because of this multiplicity, we used our 72 identified clumps for all counting-based analyses and our 206 identified dendrogram structures for the size-linewidth plots to reflect the full range of the GMC size scales.

Both dendrogram and clump-finding algorithms were run on the non-primary beam-corrected data cube for the $\mathrm{^{12}CO(2-1)}$ emission to better identify the boundaries of the structures. Once the structure boundaries were assigned, we then applied the same pixel assignments to the primary beam-corrected data cubes for all three of the CO emission lines.

\subsection{Clumps}\label{sec:clumps}
To separate the emission into non-overlapping clumps, we used the {\tt\string quickclump} \citep{2017ascl.soft04006S} algorithm with the added $\mathrm{I_{minpk}}$ parameter which requires valid clumps to have a minimum peak intensity (\url{https://github.com/indebetouw/quickclump}). This allows for the elimination of clumps with faint peaks while still including the faint emission within valid clumps. We ran the algorithm with the following parameters: {\tt\string Tcutoff}=$4.3\sigma$, {\tt\string Iminpk}=$5.75\sigma$, {\tt\string dTleaf}=$1.5\sigma$,  {\tt\string Npixmin}=300, and {\tt\string Nlevels}=1000 where $\sigma$ is the $\sigma_{rms}$ value for $\mathrm{^{12}CO(2-1)}$ shown in Table \ref{tab:obslog}. This limited the algorithm to include only emission above $4.3\sigma$ and find clumps with local maxima greater than $5.75\sigma$ that are at least $1.5\sigma$ apart. Additionally, we require a clump to contain 300 pixels ($\sim$ 1.7 beam areas). Any \textcolor{black}{local \textcolor{black}{maximum}} that does not meet those requirements was merged with adjacent valid clumps or deleted if there are no clumps nearby to merge with. These parameters were optimized to include the maximum amount of emission while still excluding clumps that do not appear in the other emission lines. Additionally, these parameters result in sizes that are in general agreement with previously measured clouds such as the Firecracker \citep{2015ApJ...806...35J,2019ApJ...874..120F} and the cloud around the SSC B1 \citep{2021PASJ...73S..35T}. This resulted in a total of 72 individual clumps \textcolor{black}{that contain 91 \% of the total cube emission.} 

Without requiring the conditions mentioned above, the parameters do not necessarily need to be as finely tuned as they are. Setting the cutoff, minimum intensity, and
minimum separation all to somewhere around $5\sigma$ would give similar overall GMC properties, though the cloud identification would not be as physically motivated.

\subsection{Dendrogram Structures}\label{sec:dendro}
We used the package {\tt\string astrodendro} \citep{Rosolowsky_2008} to decompose the emission into hierarchical structures which are separated into three categories: leaves, branches, and trunks. Leaves are the most isolated category, which represents clouds with no resolved substructure. Branches are larger structures than leaves, which both have substructures as well as being smaller substructures of larger cloud structures called trunks. Trunks are the largest structures and are not substructures of any other cloud. 
Isolated clouds that neither have substructure nor are part of larger structures can be categorized as both trunks and leaves, however, they are sorted into the leaf category by convention.

\begin{figure*}
\centering
\includegraphics[width=\textwidth]{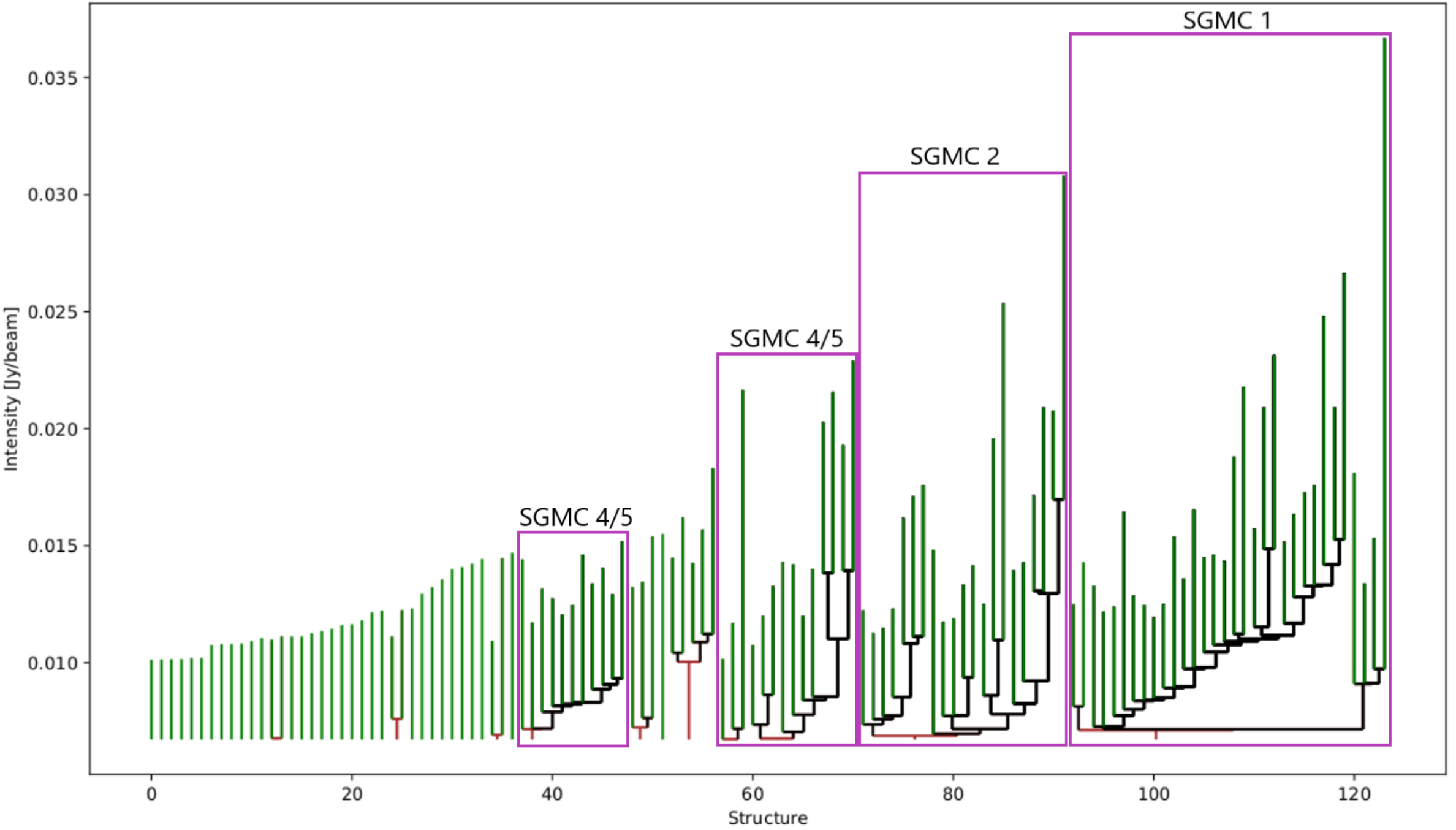}
\caption{Dendrogram for the overlap region created by {\tt\string astrodendro} where each vertical line is a structure. The red lines are trunks, the black lines are branches, and the green lines are leaves. The structures corresponding to SGMCs are boxed in magenta and labeled accordingly.
\label{fig:1}}
\end{figure*}

\begin{figure}
    \centering
    \includegraphics[width=.45\textwidth]{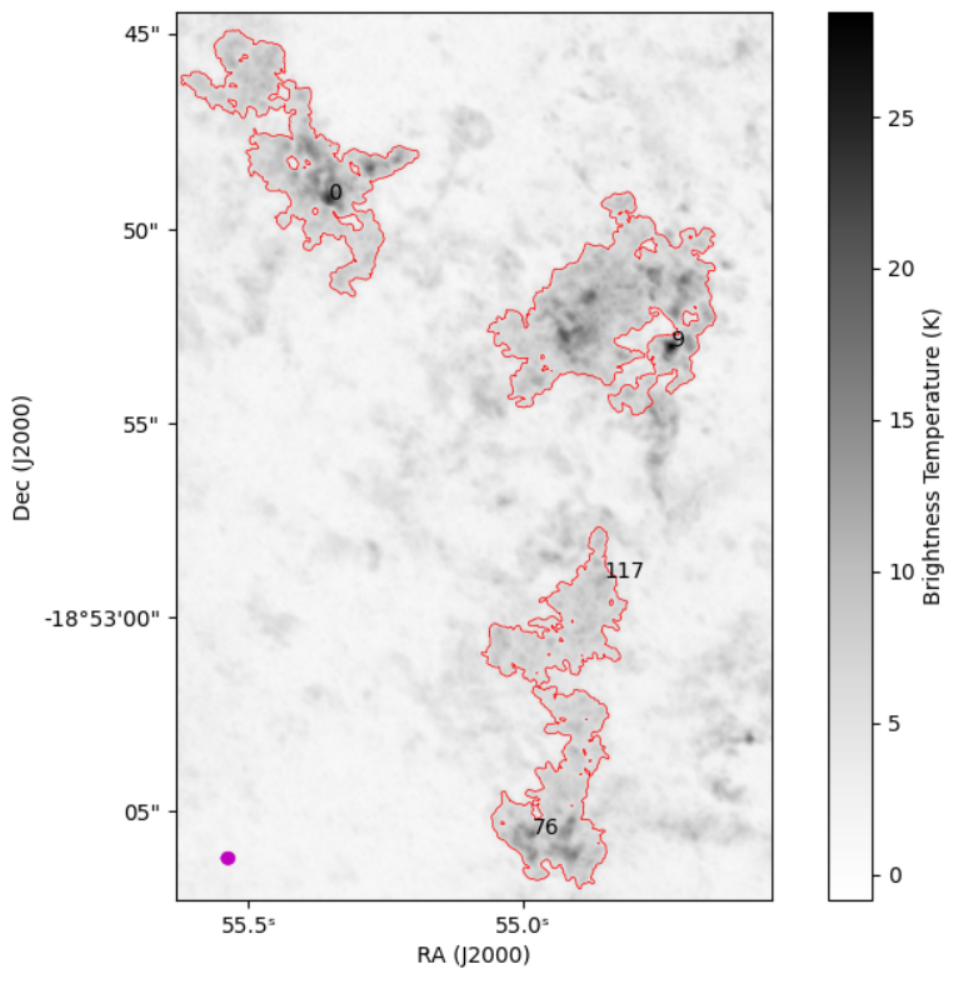}
    \caption{Dendrogram trunks associated with SGMCs 1,2, and 4/5 plotted over the $3\sigma$ peak intensity (moment 8) map of the $\mathrm{^{12}CO(2-1)}$ emission. The structure number is located at the point of maximum intensity (CO max) of each structure and the synthesized beam is shown in magenta in the bottom left corner.}
    \label{fig:trunks}
\end{figure}

The structures identified in the Antennae overlap region were found using the following parameters: {\tt\string min\_value}$=3\sigma$, {\tt\string min\_delta}$=1.5\sigma$, {\tt\string min\_peak}=$4.5\sigma$, and {\tt\string min\_npix}=2 beam areas. This resulted in a total of 206 structures with 124 leaves, 72 branches \textcolor{black}{(no correlation to the 72 identified clumps)}, and 10 trunks. These structures are visualized in the dendrogram in Figure \ref{fig:1}. The dendrogram shows the Antennae overlap having a few large main structures with a lot of substructures along with some smaller isolated structures with little to no substructures. These large structures generally align with the super giant molecular cloud (SGMC) classifications from \citet{2000ApJ...542..120W} which used 3.15" x 4.91" observations from the Caltech Millimeter Array. \textcolor{black}{The} SGMCs are made up of four dendrogram trunks with the structure IDs (ncls) 0, 9, 76, and 117 in the GMC catalogs (Appendix \ref{sec:catalog}) which make up SGMCs 1, 2, and 4/5 respectively. \textcolor{black}{The} structures are plotted on top of the peak intensity (moment 8) map of the $\mathrm{^{12}CO(2-1)}$ emission in Figure \ref{fig:trunks}. These structures that correspond to SGMCs are also labeled in Figure \ref{fig:1}. Together, the trunks associated with these SGMCs contain 96\% of the total dendrogram-captured mass and 88\% of the total area. 

\section{Calculating Physical Properties}\label{sec:props}
Using the primary beam corrected data cubes, we calculated the radius, linewidth, mass, and virial parameter, $\mathrm{\alpha_{vir}}$, for each molecular cloud identified by the {\tt\string quickclump} and {\tt\string astrodendro} algorithms. The properties calculated in this section are summarized in Tables \ref{tab:12ClumpProps} - \ref{tab:13DendroProps} and are used in the analysis in Sections \ref{sec:antprops} and \ref{sec:companalasys}.

\subsection{Sizes and Linewidths}\label{sec:calcslw} 
The radius of each clump was found by fitting an ellipse to the half-width at half-max (HWHM) of the clump. The fitted major and minor axes of the ellipse were then converted into the radius of a circle with the same area 
\begin{equation}\label{eq:rad}
R_{HWHM}=\sqrt{\sigma_{maj}\sigma_{min}}. 
\end{equation}
The error in $\mathrm{R_{HWHM}}$, $\mathrm{\delta{R}}$, was then estimated based on the circularity of the fitted ellipse. Both $\mathrm{R_{HWHM}}$ and $\mathrm{\delta{R}}$ were converted to from HWHMs to the $\sigma$ of a Gaussian such that $\mathrm{2R_{HWHM}=2.35\sigma_{R}}$. The velocity dispersion and its error, $\mathrm{\sigma_{v}}$ and $\mathrm{\delta_{v}}$, were found by fitting  distributions ({\tt\string scipy\text{.}optimize\text{.}curve\_fit}\textcolor{black}{,} \citet{2020NatMe..17..261V}) to the intensity-weighted mean line profiles. From the fitted Gaussian, $\mathrm{\sigma_{v}}$ was taken from the width (FWHM = 2.35$\mathrm{\sigma_{v}}$) and $\delta_{v}$ from the corresponding value in the covariance matrix. R and $\mathrm{\sigma_{v}}$ and their associated errors were then deconvolved by using Equations \eqref{eq:deconv} and \eqref{eq:veldeconv} respectively.

\begin{equation}\label{eq:deconv}
\sigma_{R,dc}=\sqrt{R^{2}-(\theta_{beam}/2)^{2}},
\end{equation}
\begin{equation}\label{eq:veldeconv}
\sigma_{v, dc}=\sqrt{\sigma_{v}^{2}-(\Delta_{v}/2.35)^{2}},
\end{equation}

Here, $\mathrm{\theta_{beam}}$ is the beam size in pc and $\Delta_{v}$ is the velocity resolution of 5 $\mathrm{km\:s^{-1}}$. The values for $\mathrm{\sigma_{R,dc}}$ and its error were additionally scaled by $\mathrm{\eta}$ to represent the effective radius R. We adopt $\mathrm{\eta}$ = 1.91 to maintain consistent size definitions with earlier studies \citep{1987ApJ...319..730S,2007ApJ...654..240R,Miura_2018,10.1093/mnras/stab085}\footnote{\textcolor{black}{\citet{10.1093/mnras/stab085} uses $\mathrm{\eta=1.18}$ in the originally published datasets, however, we have rescaled the radii to use our $\mathrm{\eta=1.91}$ for consistency.}}. For the rest of this paper,  $\mathrm{\sigma_{v}}$ and R refer to the deconvolved velocity dispersions and effective radii. 

The exact area, $\mathrm{A_{exact}}$, of each cloud was also calculated from the total number of pixels within a clump or dendrogram structure. These areas were on average $2.30 \pm 0.33$ times the area calculated using R ($\mathrm{A_{R}}$), \textcolor{black}{and would result in exact area-calculated R values $\mathrm{\sim1.5}$ times the fitted Rs}. The fitted Rs of GMCs with $\mathrm{A_{exact}/A_{R}}$ values significantly larger ($\mathrm{>}$ 5 times the standard deviations of the average $\mathrm{A_{exact}/A_{R}}$ ratios) were discarded as poor fits that did not accurately represent the sizes of the clouds.

\subsection{Mass}\label{sec:mass}
Mass estimates vary significantly across different methods of calculation. The LTE (local thermodynamic equilibrium) method uses local thermodynamic equilibrium assumptions to obtain CO excitation temperatures, optical depths, and column densities. When  LTE conditions apply, this method is able to accurately estimate the masses of individual GMCs that have varying temperatures. However, the LTE method requires multiple lines of strong emission and becomes less accurate when either of the lines does not have strong emission, the assumed optically thick line becomes optically thin, or if the lines' level populations are not well described by a Boltzmann distribution. Additionally, the LTE method is heavily dependent on abundance ratios. Regardless of how accurate the calculated column density is, poorly constrained abundance ratios can introduce a large amount of error\textcolor{black}{,} as seen in Figure \ref{fig:XCOLTE}.

\textcolor{black}{In contrast, using a constant $\mathrm{CO\text{-}to\text{-}H_{2}}$ factor to calculate mass works best at large size scales which reflect the average physical conditions. As the size scales become smaller, the $\mathrm{CO\text{-}to\text{-}H_{2}}$ factor is expected to have large variations along a line-of-sight \citep{2013ARA&A..51..207B}.} However, the CO-to-$\mathrm{H_{2}}$ method only requires data for one emission line and is not as dependent on abundance ratios as the LTE method is. This makes using a constant $\mathrm{CO\text{-}to\text{-}H_{2}}$ factor much more accessible and thus is more widely used. The lack of additional variables, such as abundance ratios, also allows for a simpler comparison between different data sets since the only factors in determining the masses are the $\mathrm{CO\text{-}to\text{-}H_{2}}$ factor and the observed CO luminosity.

Because these two methods can differ significantly from one another, we calculated the cloud masses using both methods.

\subsubsection{LTE}\label{sec:ltemass}
The LTE masses for the clouds were calculated using the $\mathrm{^{12}CO(2-1)}$ and $\mathrm{^{13}CO(2-1)}$ lines\textcolor{black}{,} which we assumed to be optically thick and optically thin respectively. These lines are also assumed to have the same excitation temperature, $\mathrm{T_{x}}$, which does not vary across the line profile. This is generally described by
\begin{equation}\label{eq:1}
T_{b} =T_{ul}(1-e^{-\tau_{\nu}}) \left [ \frac{1}{e^{\frac{T_{ul}}{T_{x}}}-1}-\frac{1}{e^{\frac{T_{ul}}{T_{bg}}}-1}   \right ], 
\end{equation}
where $\mathrm{T_{b}}$ is the brightness temperature and $\mathrm{\tau_{\nu}}$ is the optical depth. Because $\mathrm{^{12}CO}$ is assumed to be optically thick ($\mathrm{e^{-\tau_{\nu}}\approx0}$), the optical depth term $\mathrm{1-e^{-\tau_{\nu}}}$ approximates to 1 and Equation \eqref{eq:1} can be rearranged to solve for $\mathrm{T_{x}}$
\begin{equation}\label{eq:2}
T_{x}=T_{ul} \left [ ln\left (\frac{-T_{12}e^{\frac{T_{ul,12}}{T_{bg}}}+T_{12}-T_{ul}e^{\frac{T_{ul}}{T_{bg}}}}{\textcolor{black}{-}T_{ul,12}-T_{12}e^{\frac{T_{ul,12}}{T_{bg}}}+T_{12}} \right )  \right ]^{-1}.
\end{equation}

 Here $\mathrm{T_{12}}$ is the brightness temperature for $\mathrm{^{12}CO(2-1)}$, $\mathrm{{T_{ul, 12}}=11.07}$ K and $\mathrm{T_{bg}=2.73}$ K. Once $\mathrm{T_{x}}$ is calculated, it can then be inserted back into Equation \eqref{eq:1} for $\mathrm{^{13}CO(2-1)}$. Since $\mathrm{^{13}CO(2-1)}$ is optically thin, Equation \eqref{eq:1} can be rearranged to calculate $\tau_{\nu,13}$

\begin{equation}\label{eq:3}
\tau_{\nu, 13} = -ln\left [ 1-\frac{T_{13}}{T_{ul, 13}}\left [ \frac{1}{e^\frac{T_{ul, 13}}{T_{x}}-1}-\frac{1}{e^\frac{T_{ul, 13}}{T_{bg}}-1}\right ]^{-1} \right ],
\end{equation}
using $\mathrm{T_{ul,13}=10.58}$ K. The column density can then be found using the Equation 101 from \citet{2015PASP..127..266M}
\begin{equation}\label{eq:4}
N_{tot} = \frac{8\pi\nu_{0}^{2}Q}{c^{2}A_{ul}g_{u}}\frac{1}{1-e^{\frac{-T_{ul,13}}{T_{x}}}}\int \tau_{\nu}d\nu.
\end{equation}

After using $\mathrm{Q = \frac{T_{x}}{B_{0}}+\frac{1}{3}}$, $\mathrm{B_{0}=2.664}$ K, and $\mathrm{\tau_{\nu,13}}$, Equation \eqref{eq:4} becomes

\begin{equation}\label{eq:5}
\frac{N^{13}}{cm^{-2}} = \frac{8\pi\nu_{0}^{2}}{c^{2}A_{ul}g_{u}}\left ( \frac{T_{x}}{2.644}+\frac{1}{3} \right )\frac{1}{1-e^{\frac{-T_{ul,13}}{T_{x}}}}\int \tau_{\nu,13}d\nu.
\end{equation}
Using Equations \eqref{eq:2}, \eqref{eq:3}, and \eqref{eq:5}, we determined the column density of $^{13}$CO(2-1) per pixel for each clump. We then calculated the total molecular mass of each GMC with 
\begin{equation}\label{eq:ltemass}
    M_{LTE}=1.36m_{H_{2}}\frac{H_{2}}{^{13}CO}A_{pix}\sum_{i} N_{i}^{13},
\end{equation}
where $\mathrm{1.36m_{H_{2}}}$ is the mean particle mass, $\mathrm{H_{2}/^{13}CO}$ is the abundance ratio of $\mathrm{H_{2}}$ to $\mathrm{^{13}CO}$ molecules, $\mathrm{A_{pix}}$ is the area of a pixel in $\mathrm{pc^{2}}$, and $\mathrm{\sum_{i}N_{i}^{13}}$ is the $\mathrm{^{13}CO}$ column density calculated in Equation \eqref{eq:5} summed over all pixels $\mathrm{i}$ in each cloud.

We used abundance ratios of $\mathrm{H_{2}/^{13}CO=1.4\times 10^{6}}$ by combining $\mathrm{^{12}CO/^{13}CO}=70$ and $\mathrm{H_{2}/^{12}CO=2\times 10^{4}}$, although the abundance ratio could be anywhere in the range $\mathrm{H_{2}/^{13}CO=4\times 10^{5}-2\times 10^{7}}$. This is because both the $\mathrm{^{12}CO/^{13}CO}$ and $\mathrm{H_{2}/^{12}CO}$ ratios have wide ranges of possible values. The overlap region of the Antennae galaxies has been recorded to have $\mathrm{^{12}CO/^{13}CO}=40-200$ with 70 being the most frequently found value \citep{Zhu_2003}. Although the $\mathrm{H_{2}/^{12}CO}$ ratio has not been extensively studied in the Antennae overlap, \citet{2014A&A...563A..97G} found that the abundance ratio ranges between $10^{4}-10^{5}$ for the earliest stage of massive star formation using several sources in the Milky Way galactic plane. This stage encompasses most of the GMCs in the Antennae overlap\textcolor{black}{,} which have not yet formed massive protostellar objects.

Using the method and abundance ratios described above gives the clump masses, as well as mean excitation temperatures and optical depths \textcolor{black}{that are} described in Table \ref{tab:ltemass stats}.

\begin{deluxetable}{cccc}
\tablecaption{LTE Mass Statistics \label{tab:ltemass stats}}
\tablehead{\colhead{} & \colhead{Mass} & \colhead{$\mathrm{T_{x}}$} & \colhead{$\mathrm{\tau_{\nu,13}}$} \\ \colhead{} &\colhead{$\mathrm{M_{\odot}}$} & \colhead{K} & \colhead{}}
\startdata
Minimum &$5.95\times10^{4}$&13&0.04\\
Maximum &$1.14\times10^{7}$&25&0.22\\
Mean &$1.53\times10^{6}$&17&0.11\\
Total &$1.10\times10^{8}$&...&...\\
\enddata
\end{deluxetable}

\subsubsection{$CO \text{-} to \text{-} H_{2}$ Conversion Factor}
Because the LTE method requires multiple lines to be traced, it is not as reliable for areas with weaker $\mathrm{^{13}CO}$ emission. Due to the requirement of two strong emission lines, many GMC catalogs calculate mass using a $\mathrm{CO\text{-}to\text{-}H_{2}}$ conversion factor \textcolor{black}{that} directly converts CO luminosity to the column density of $\mathrm{H_{2}}$. Although using a standard CO-to-$\mathrm{H_{2}}$ factor is simple to implement, it is difficult to calculate accurate CO-to-$\mathrm{H_{2}}$ factors due to their dependence on the physical conditions of a GMC. This is especially true for galaxies with non-typical metallicities and excitation conditions that are not captured with CO-to-$\mathrm{H_{2}}$ factors calibrated to Milky Way molecular clouds \citep{2013ARA&A..51..207B}.

Although the CO-to-$\mathrm{H_{2}}$ can be heavily reliant on assumptions or previous mass calculations, it is widely used to calculate mass in a lot of GMC catalogs. In order to be able to directly compare the properties of the Antennae overlap GMCs to other GMC catalogs that use this method, we also calculated cloud mass using a constant $\mathrm{CO\text{-}to\text{-}H_{2}}$. To do so we first calculated the integrated flux density, $\mathrm{L_{CO}}$ given by

\begin{equation}
    L_{CO}=A_{pix}R_{l}\sum_{i} T_{i}\Delta v \;\;  \mathrm{K\:km\:s^{-1}\:pc^{2}},
\end{equation}
where $\mathrm{R_{l}}$ is the line ratio from the measured emission line to $\mathrm{^{12}CO(1-0)}$ and $\mathrm{\sum_{i}T_{i}\Delta v}$ is the brightness temperature in K $\mathrm{km^{-1}}$ summed over all pixels $\mathrm{i>5\sigma_{rms}}$ and the velocity channel width, $\mathrm{\Delta v}$, in the cloud. Each of the $\mathrm{R_{l}}$ ratios were taken from \citet{Zhu_2003} \textcolor{black}{with typical line intensity for $\mathrm{CO(1-0)/CO(2-1)}$ and $\mathrm{CO(1-0)/^{13}CO(2-1)}$ of 0.95 and 22 respectively.}

The luminosity is then converted into mass using the standard $\mathrm{X_{CO}=0.5\times10^{20}\;cm^{-2}(K\:km\:s^{-1})^{-1}}$ for starburst galaxies \citep{2013ARA&A..51..207B}, as well as the same ratio of $\mathrm{H_{2}}$ mass to total mass used for the LTE mass calculations.

\begin{equation}
    M_{lum} = 1.36m_{H_{2}}X_{CO}L_{CO}.
\end{equation}

\begin{figure}
\includegraphics[scale=.5]{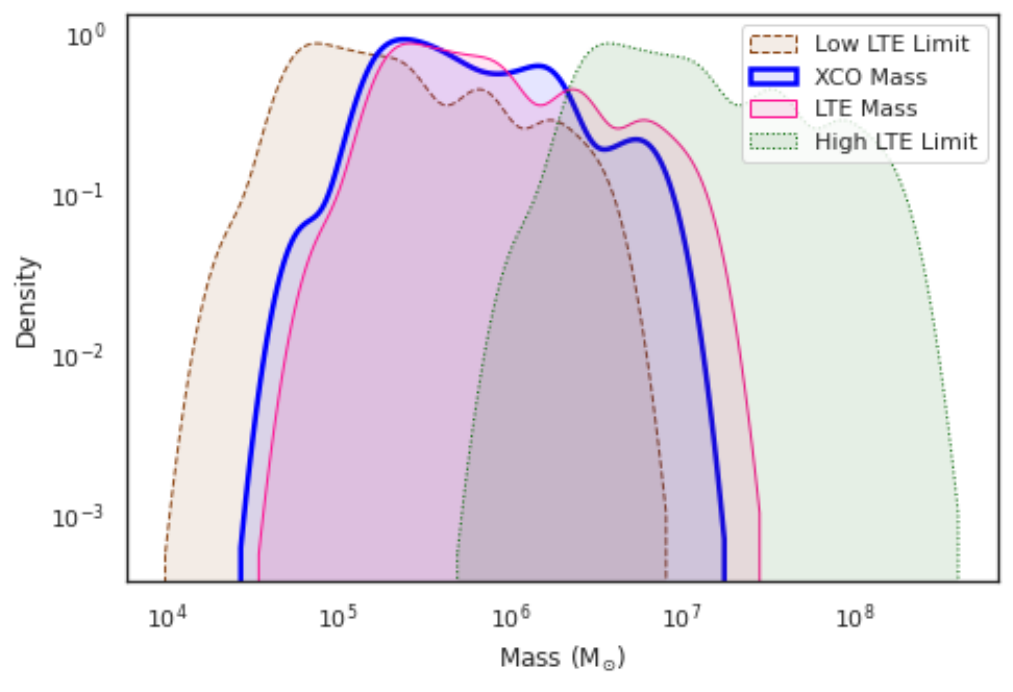}
\caption{Distributions of mass calculated by both the LTE and $\mathrm{CO\text{-}to\text{-}H_{2}}$ methods represented by KDEs. The low and high LTE limits are defined using the lowest and highest possible abundance ratios of  $\mathrm{H_{2}/^{13}CO=4\times 10^{5}}$ and $\mathrm{H_{2}/^{13}CO=2\times 10^{7}}$ respectively. The distribution simply referred to as the LTE mass uses the $\mathrm{H_{2}/^{13}CO=1.4\times 10^{6}}$ ratio which is used for all analyses with an LTE-derived mass. The $\mathrm{X_{CO}}$ mass uses a standard starburst $\mathrm{CO\text{-}to\text{-}H_{2}}$ factor of $\mathrm{X_{CO}}$ = $\mathrm{0.5\times10^{20}\;cm^{-2}(K\:km\:s^{-1})^{-1}}$.\label{fig:XCOLTE}}
\end{figure}

This gives $\mathrm{X_{CO}}$ masses with similar distributions to the LTE masses calculated using $\mathrm{H_{2}/^{13}CO=1.4\times 10^{6}}$ as shown in Figure \ref{fig:XCOLTE}. $\mathrm{M_{lum}}$ is also in the middle of the upper and lower LTE mass limits, although it is slightly closer to the lower limit. These distributions are represented as kernel density estimates (KDEs) which estimate a continuous probability function using kernel smoothing. All of the KDE plots in this paper used bandwidths of 0.25 dex.

We found that these two methods of mass calculation presented here produced general agreement within a factor of 1.34 for the best estimate and 10.1 for the worst. The LTE method tended to produce slightly lower masses, which we infer as a result of different possible abundance ratios. We also note that the assumption of LTE might not be valid. If $\mathrm{^{12}CO}$ and $\mathrm{^{13}CO}$ are not thermalized \textcolor{black}{and} $\mathrm{T_{x,13}}\:<$ $\mathrm{T_{x,12}}$, these masses are lower estimates. On the other hand, if $\mathrm{^{12}CO}$ is not actually optically thick then these masses are upper estimates.

Using the best-fit abundance ratio of $\mathrm{H_{2}/^{13}CO}$ $=1.4\times 10^{6}$ for the LTE-derived mass, we determined that both methods produce cloud masses that could be drawn from the same distribution as the other. This was done using a Kolmogorov–Smirnov (K-S) test which gives a  K-S statistic as a value between 0 and 1 representing the distance between the two distributions, as well as providing p-values to determine the likelihood of the data sets being drawn from the same distribution. The K-S statistic and p-value between the two mass distributions are 0.148 and 0.524 respectively, indicating that they could be drawn from the same distribution. This indicates that the $\mathrm{CO\text{-}to\text{-}H_{2}}$ mass is \textcolor{black}{consistent with LTE mass calculated when using the best-fit ratios mentioned above.}

\subsection{Composite Properties}\label{sec:calcvir}
Using the R, $\mathrm{\sigma_{v}}$, and $\mathrm{M_{lum}}$ calculated as shown above, we calculated each cloud's virial parameter $\mathrm{\alpha_{vir}}$, external pressure $\mathrm{P_{e}}$, surface density $\mathrm{\Sigma}$, and free-fall time $\mathrm{t_{ff}}$. 

The virial parameter $\mathrm{\alpha_{vir}}$ measures the balance between the gravitational and kinetic energy in a molecular cloud such that $\mathrm{\alpha_{vir}=1}$ when the cloud is in virial equilibrium and is calculated by
\begin{equation}\label{eq:alphavir}
\alpha_{vir}=\frac{5\sigma_{v}^{2}R}{GM}
\end{equation}
with $G$ as the gravitational constant. Subvirial GMCs with $\mathrm{\alpha_{vir}<1}$ are dominated by potential energy and therefore are likely to collapse while supervirial GMCs with $\mathrm{\alpha_{vir}>1}$ are dominated by kinetic energy. Supervirial GMCs are not bound by gravity and thus could disperse unless they are bound by an external pressure. Assuming that our observed GMCs are not dispersing, we calculated the external pressure that is needed for them to remain bound with Equation 3 from \citet{1989ApJ...338..178E}

\begin{equation}\label{eq:pressure}
P_{e}=\frac{3\prod M\sigma_{v}^{2}k}{4\pi R^{3}}
\end{equation}
where we adopt $\mathrm{\prod=0.5}$ \citep{2015ApJ...806...35J} as the ratio between the local and mean ionized gas densities and $k$ is the Boltzmann constant. Both surface density $\mathrm{\Sigma}$ and free-fall time $\mathrm{t_{ff}}$ were calculated using only each GMCs' mass M and radius R

\begin{equation}\label{eq:density}
    \Sigma = \frac{M}{\pi R^{2}}
\end{equation}

\begin{equation}\label{eq:tff}
    t_{ff} = \sqrt{\frac{3\pi}{32G\rho}} = \sqrt{\frac{\pi^{2}R^{3}}{4GM}}.
\end{equation}

As mentioned in Section \ref{sec:calcslw}, cloud radii calculated from the exact area are on average \textcolor{black}{1.5} times larger than those calculated from fitting ellipses to the clouds' HWHMs. Using those larger estimations of R from $\mathrm{A_{exact}}$ to calculate the properties above would result in larger virial parameters and free fall times while decreasing the pressures and surface densities.

\section{Pre-SSC-Forming Clouds}\label{sec:antprops}
\begin{figure*}
    \centering
    \includegraphics[width=.97\textwidth]{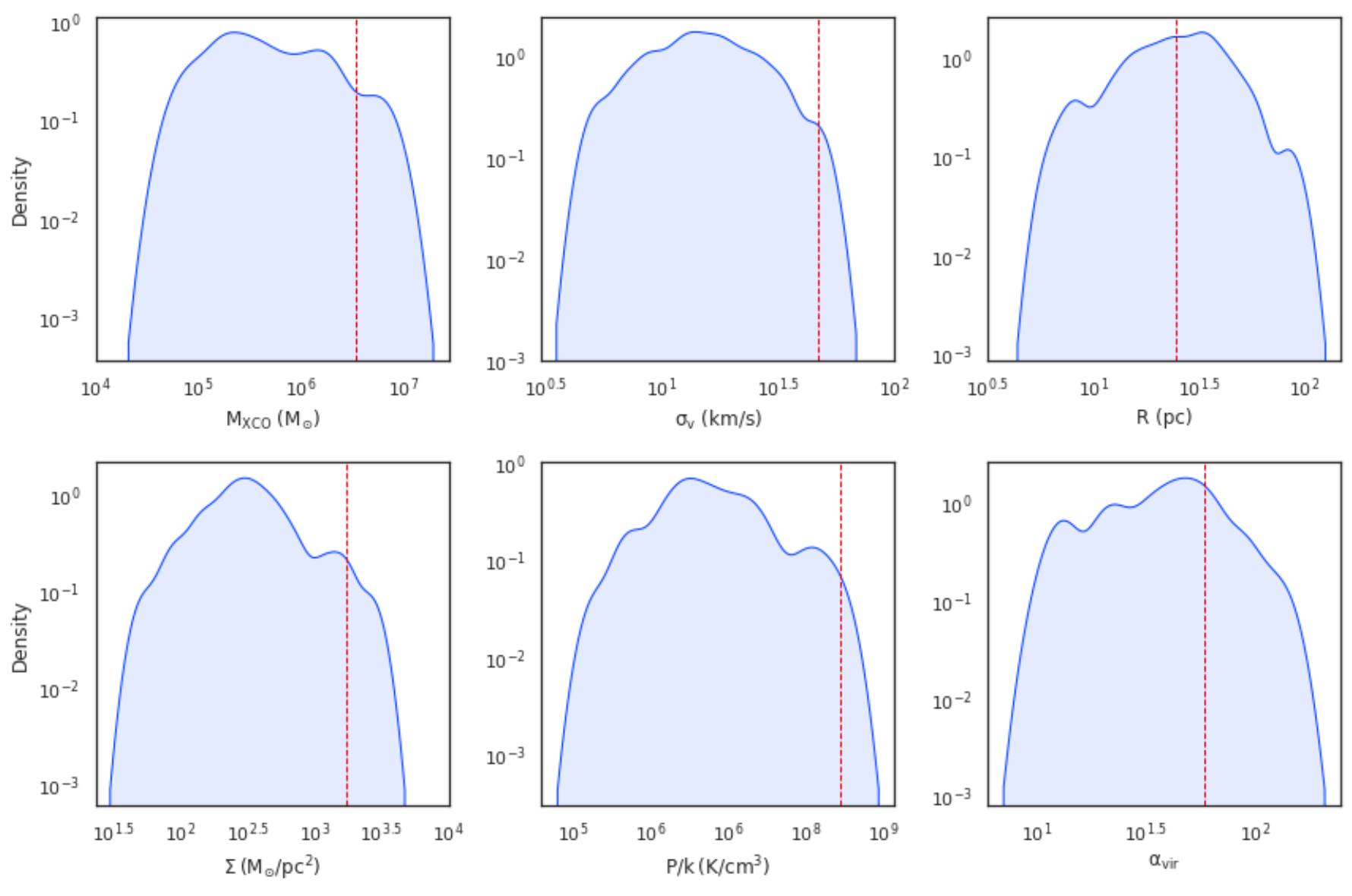}
    \caption{Physical property distributions of \textcolor{black}{clumps in the} Antennae overlap represented as KDEs with smoothing parameters of 0.25 dex. The mass-based properties are calculated with $\mathrm{M_{lum}}$. The dotted red line shows where the Firecracker falls on each of the distributions.}
    \label{fig:together}
\end{figure*}

The properties calculated in Section \ref{sec:props} for the Antennae overlap are summarized in Table \ref{tab:stats}, and the property distributions represented as KDEs are shown in Figure \ref{fig:together}. 
\begin{deluxetable}{ccccc}
\tablecaption{Clump Property Statistics\label{tab:stats}}
\tablehead{\colhead{Property} & \colhead{Min} & \colhead{Max} & \colhead{Mean} & \colhead{$\mathrm{\sigma}$}} 
\startdata
R & 6.51 & 84.9 & 27.7 & 14.2 \\
$\mathrm{\sigma_{v}}$ ($\mathrm{km\:s^{-1}}$) & 5.24 & 46.6 & 17.9 & 9.04 \\
$\mathrm{M_{lum}}$ ($\mathrm{10^{5}\:M_{\odot}}$)& 0.63 & 77 & 11.4 & 15.8 \\
$\mathrm{\Sigma_{X_{CO}}}$ ($\mathrm{M_{\odot}\:pc^{-2}}$)& 53.9 & 2650 & 441 & 452 \\
$\mathrm{P_{e, X_{CO}}/k}$ ($\mathrm{10^{7}Kcm^{-3}}$)& 0.02 & 28.1 & 2.11 & 4.74 \\
$\mathrm{\alpha_{vir,X_{CO}}}$ & 3.45 & 42.4 & 13.8 & 7.84 \\
$\mathrm{t_{ff,X_{CO}}}$ ($\mathrm{10^{-2}Myr}$)& 1.05 & 7.39 & 3.26 & 1.22 \\
$\mathrm{M_{LTE}}$ ($\mathrm{10^{5}\:M_{\odot}}$)& 0.86 & 114 & 17.0 & 25.5 \\
$\mathrm{\Sigma_{LTE}}$ ($\mathrm{M_{\odot}\:pc^{-2}}$) & 55.6 & 5830 & 606 & 789 \\
$\mathrm{P_{e,LTE}/k}$ ($\mathrm{10^{7}Kcm^{-3}}$)& 0.04 & 3.20 & 24.1 & 51.3 \\
$\mathrm{\alpha_{vir,LTE}}$ & 1.62 & 42.8 & 12.2 & 9.60 \\
$\mathrm{t_{ff,LTE}}$ ($\mathrm{10^{-2}Myr})$& 0.71 & 7.28 & 2.89 & 1.17 \\
\enddata
\tablecomments{Statistics of selected physical properties for the $\mathrm{^{12}CO(2-1)}$ emission in all 72 clumps.} 
\end{deluxetable}

Figure \ref{fig:together} shows that each of the 6 properties has an approximately normal distribution. This is confirmed by a Shapiro-Wilk test for normality \citep{SW}, which gives p-values $\mathrm{>>}$ 0.05 for all of the distributions. Although the distributions are generally normal, each of the $\mathrm{M_{lum}}$, $\mathrm{\sigma_{v}}$, $\mathrm{\Sigma}$, and $\mathrm{P_{e}/k}$ KDEs show a feature on the high end of the distribution of which the Firecracker is a part of. The Firecracker (GMC 1 in all tables) was first identified as a proto-SSC by \citet{2014ApJ...795..156W}. Several follow-up analyses constrained its physical properties and associated thermal radio emission indicative of already-formed stars \citep{2015ApJ...806...35J,2019ApJ...874..120F}. So far, it has been the only example found of a cloud in the earliest stage of star formation ($\mathrm{\sim\:1\:Myr}$) that has the potential to form an SSC that does not have any significant thermal radio emission.

In order to determine if there are other clouds that may be in a similar stage of SSC formation, we looked for other objects with particularly extreme physical properties (\ref{sec:sscprops}) and analyzed their thermal radio emission associated with star formation (\ref{sec:ionizingflux}.)

\subsection{GMCs with Extreme Physical Properties\label{sec:sscprops}}

GMCs in the earliest stages of SSC formation are expected to have a large amount of mass and to be among the most massive and dense clouds since the gas has not been disrupted by significant star formation \citep{2015ApJ...801...25L,2015ApJ...806...35J,2018ApJ...853..125J,2019ApJ...874..120F,2022ApJ...928...57H}. They also require high external pressures to remain bound \citep{Elmegreen_2001,2019ApJ...874..120F}. In order to identify potential pre-SSC-forming clouds, we used the {\tt\string papanda} implementation of the Generalized Extreme Studentized Deviate (GESD) test \citep{1983Rosner} to detect outliers in each of the calculated physical properties with an upper bound of 20 outliers. Particular interest was given to GMCs with stand-out pressures, densities, and virial parameters, since they are less dependent on the size of the cloud.

\begin{figure*}
    \centering
    \includegraphics[width=0.99\textwidth]{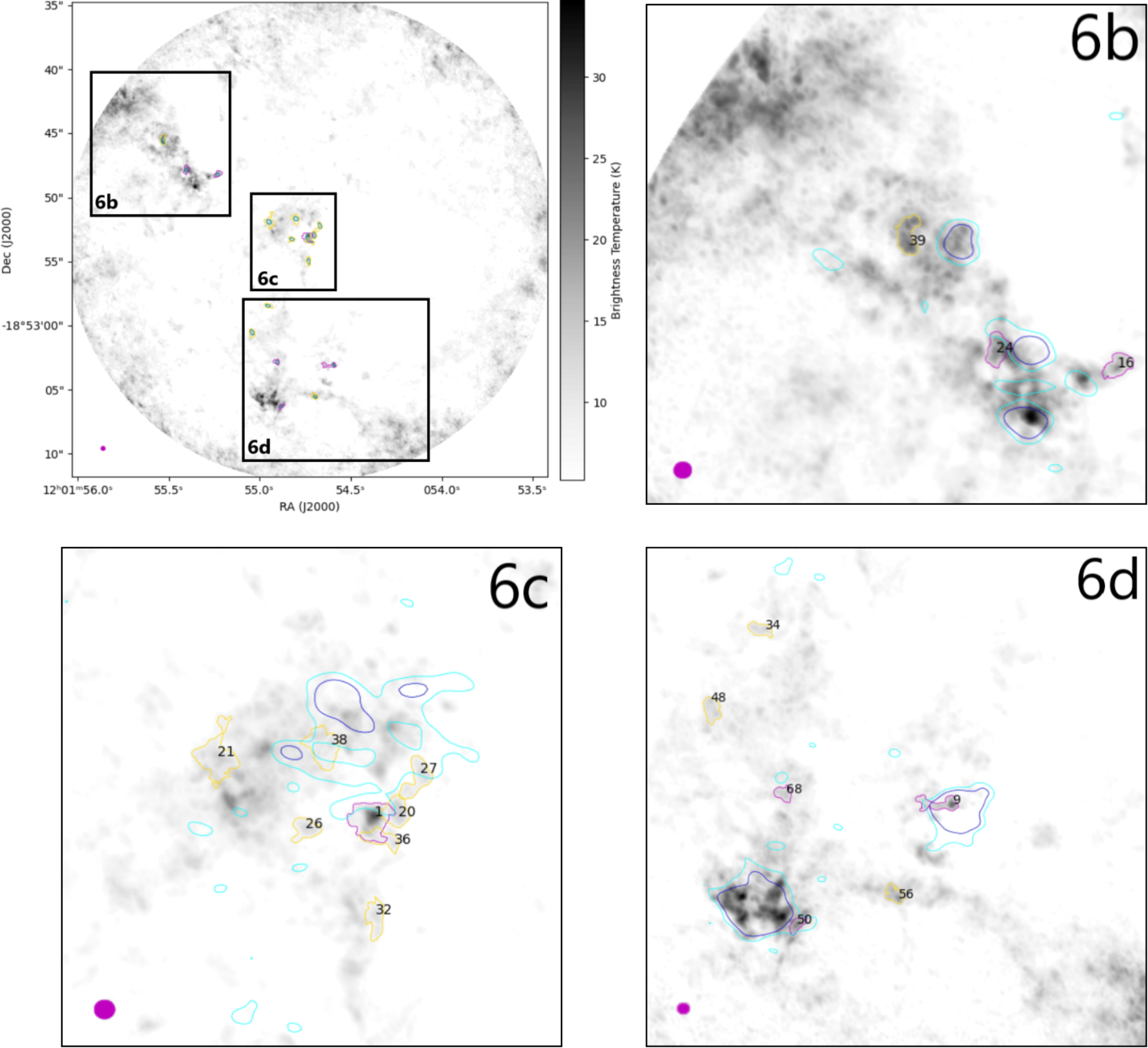}
    \caption{Contours of clumps that have or are close to having the necessary conditions to form SSCs plotted over top of the peak luminosity (moment 8) map of the $\mathrm{^{12}CO(2-1)}$ emission data. Clumps identified as outliers are outlined in magenta while the others are outlined in gold. Clump ID numbers are plotted at each corresponding local maximum intensity and the synthesized beam is shown at the bottom left in magenta. The full view of the Antennae overlap is shown on the top left (Figure 6a) with each of the three SGMC structures boxed in black. The zoomed-in view of SGMCs 1, 2, and 4/5 are shown in Figures 6b-6d respectively. The zoomed-in figures also include the 3$\sigma$ (cyan) and 5$\sigma$ (dark blue) contours of the 3.6 cm ionizing flux data.}\label{fig:sscforming}
\end{figure*}

Outliers were detected at the 0.05 significance level, meaning that there is a 95\% chance the clump is an outlier. Clumps that are identified as outliers in at least two of the three composite properties (density, pressure, and virial parameter) are bolded in Table \ref{tab:sscprops} and outlined in magenta in Figure \ref{fig:sscforming}.

All 6 of these clumps have $\mathrm{\Sigma>10^{3}\:M_{\odot}\:pc^{-2}}$, $\mathrm{P_{e}/k>10^{7}\:K\:cm^{-3}}$, and $\mathrm{R<25\:pc}$, suggesting that the GMCs are could be capable of forming SSCs based off of the hypothetical SSC progenitor GMC \citep{1997ApJ...480..235E,2015ApJ...806...35J}. In addition to these 6 outlier clumps, there are eleven other clumps that are within uncertainties of having the necessary conditions to form SSCs. All 17 of the possible SSC-forming clumps are shown in Figure \ref{fig:sscforming} with some of their physical properties summarized in Table \ref{tab:sscprops}.

While some of these clumps do not meet the exact $\mathrm{M_{lum}}>10^{6}\:M_{\odot}$, $\mathrm{R\lesssim25}$ pc, and $\mathrm{P_{e}/k>10^{7}\:K\:cm^{-3}}$ requirements, they are within uncertainties of these thresholds assuming 10\% errors in observed flux and that $\mathrm{X_{CO}}$ tends to vary by a factor of 4 in starburst galaxies. These requirements are also not hard limits and are instead estimated using physical properties of a typical globular cluster and an assumed SFE of $20-50 \%$ \citet{1994ApJ...429..177H,2001AJ....122.1888A,2008ApJ...679.1272M,2015ApJ...806...35J}. A GMC that does not meet the exact requirements could still form an SSC.

\begin{deluxetable}{cccccc}
\tablecaption{Possible SSC-Forming Clumps\label{tab:sscprops}}
\tablehead{\colhead{ncl} & \colhead{$\mathrm{M_{lum}}$} & \colhead{R} & \colhead{$\mathrm{\Sigma}$} & \colhead{$\mathrm{P_{e}/k}$} & \colhead{$\mathrm{N_{Lyc}}$}\\ 
\colhead{} & \colhead{($\mathrm{10^{6}\:M_{\odot}}$)} & \colhead{(pc)} & \colhead{($\mathrm{10^{3}\:M_{\odot}pc^{-2}}$)} & \colhead{($\mathrm{10^{7}\:Kcm^{-3}}$)} & \colhead{$\mathrm{10^{49}\:s^{-1}}$} } 
\startdata
\textbf{1} & \colorbox{h_color}{3.35$\mathrm{\pm}$0.33} & \colorbox{h_color}{24.8$\mathrm{\pm}$2.5} & 1.74$\mathrm{\pm}$0.18 & \colorbox{h_color}{28.1$\mathrm{\pm}$9.0} &16.7\\
 \textbf{9}& 0.52$\mathrm{\pm}$0.05 & \colorbox{h_color}{7.89$\mathrm{\pm}$12.4} & 2.65$\mathrm{\pm}$1.61 & \colorbox{h_color}{14.5$\mathrm{\pm}$68.5}& 78.1\\
\textbf{16} & 0.97$\mathrm{\pm}$0.1 & \colorbox{h_color}{14.9$\mathrm{\pm}$1.52} & 1.39$\mathrm{\pm}$0.14 & \colorbox{h_color}{17.9$\mathrm{\pm}$6.0}&...\\
20 & 0.78$\mathrm{\pm}$0.08 & 25.9$\mathrm{\pm}$9.4 & 0.37$\mathrm{\pm}$0.04 & \colorbox{h_color}{1.47$\mathrm{\pm}$1.62}&12.9\\
21 & 0.84$\mathrm{\pm}$0.08 & \colorbox{h_color}{23.7$\mathrm{\pm}$11} & 0.48$\mathrm{\pm}$0.06 & \colorbox{h_color}{1.56$\mathrm{\pm}$2.18}&...\\
\textbf{24} & 0.9$\mathrm{\pm}$0.09 & \colorbox{h_color}{15.6$\mathrm{\pm}$2.0} & 1.17$\mathrm{\pm}$0.12 & \colorbox{h_color}{9.04$\mathrm{\pm}$3.6}&8.6\\
26 & 0.3$\mathrm{\pm}$0.03 & \colorbox{h_color}{18.4$\mathrm{\pm}$9.5} & 0.28$\mathrm{\pm}$0.04 & 0.53$\mathrm{\pm}$0.82&...\\
27 & 0.37$\mathrm{\pm}$0.04 & \colorbox{h_color}{23.9$\mathrm{\pm}$14.9} & 0.21$\mathrm{\pm}$0.03 & 0.3$\mathrm{\pm}$0.57&17.7\\
32 & 0.36$\mathrm{\pm}$0.04 & \colorbox{h_color}{19.5$\mathrm{\pm}$7.8} & 0.31$\mathrm{\pm}$0.04 & \colorbox{h_color}{1.98$\mathrm{\pm}$2.4}&...\\
34 & 0.27$\mathrm{\pm}$0.03 & \colorbox{h_color}{16.2$\mathrm{\pm}$12.2} & 0.33$\mathrm{\pm}$0.06 & 0.42$\mathrm{\pm}$0.95&...\\
36 & 0.27$\mathrm{\pm}$0.03 & \colorbox{h_color}{18.8$\mathrm{\pm}$10.7} & 0.24$\mathrm{\pm}$0.03 & \colorbox{h_color}{1.03$\mathrm{\pm}$1.75}&...\\
38 & 0.43$\mathrm{\pm}$0.04 & \colorbox{h_color}{22.4$\mathrm{\pm}$11.4} & 0.27$\mathrm{\pm}$0.03 & 0.67$\mathrm{\pm}$1.03&17.6\\
39 & \colorbox{h_color}{1.15$\mathrm{\pm}$0.12} & \colorbox{h_color}{24.3$\mathrm{\pm}$18.5} & 0.62$\mathrm{\pm}$0.09 & \colorbox{h_color}{3.68$\mathrm{\pm}$8.39}&...\\
48 & 0.45$\mathrm{\pm}$0.05 & \colorbox{h_color}{22.0$\mathrm{\pm}$8.6} & 0.3$\mathrm{\pm}$0.03 & 0.79$\mathrm{\pm}$0.93&...\\
\textbf{50} & 0.38$\mathrm{\pm}$0.04 & \colorbox{h_color}{8.57$\mathrm{\pm}$10.8} & 1.64$\mathrm{\pm}$0.74 & \colorbox{h_color}{10.6$\mathrm{\pm}$40.3}&...\\
56 & 0.31$\mathrm{\pm}$0.03 & \colorbox{h_color}{14.1$\mathrm{\pm}$5.9} & 0.49$\mathrm{\pm}$0.07 & 0.49$\mathrm{\pm}$0.61&...\\
\textbf{68} & 0.4$\mathrm{\pm}$0.04 & \colorbox{h_color}{11.0 $\mathrm{\pm}$4.0} & 1.06$\mathrm{\pm}$0.15 & \colorbox{h_color}{5.66$\mathrm{\pm}$6.3}&...\\
\enddata
\tablecomments{Select physical properties of clumps that are within uncertainties of having the necessary conditions to form SSCs. The clumps that were previously identified as extreme GMCs are bolded and the properties that meet the $\mathrm{M_{lum}}$, R, and $\mathrm{P_{e}/k}$ thresholds are highlighted in gray. $\mathrm{N_{Lyc}}$ is only shown for clumps that are detected above the $3\sigma$ threshold.}
\end{deluxetable}

Figure \ref{fig:sscforming} shows that most of the potential SSC-forming clumps are in larger structures with a particularly high number of clumps in the middle structure, SGMC 2. This could indicate that the SGMC as a whole is a more extreme environment and is forming stars at a faster rate even when compared to the other Antennae SGMCs.

These 17 clumps make up 28\% of the total number of clumps with resolved radii and contain 17\% of the total clump-identified mass. Due to many of the clumps having large errors in their radii, the margins of error in R were not taken into account when identifying the possible SSC-forming clumps shown in Figure \ref{fig:sscforming} and Table \ref{tab:sscprops}. Taking those errors into account would bring an additional 19 clumps into the possible SSC-forming range. This would result in a total of 36 out of 61 (59\%) of all resolved clumps that contain 45\% of all clump-identified mass. Only including clumps that meet the aforementioned SSC requirements without taking errors into account gives a much more conservative estimate of 3.3\% of all resolved clumps which contain 6.2\% of the total mass.
\textcolor{black}{This shows that even with the strictest of definitions, there are still several areas in the Antennae overlap that have the theoretically necessary requirements for forming SSCs.}

\subsection{Associated Star Formation \label{sec:ionizingflux}}
In order to determine which of these possible SSC-forming clumps may already have star formation affecting the gas, \textcolor{black}{we} calculated the ionizing flux of each cloud using 3.6 cm radio continuum observations. We did so using Equation 8 from \citet{2016A&A...588A.143S}
\begin{equation}
    \mathrm{\frac{N_{Lyc}}{s}= 4.771 \times 10^{42} \left(\frac{S_{\nu}}{Jy}\right)\left(\frac{T_{e}}{K}\right)^{-0.45}\left(\frac{\nu}{GHz}\right)^{0.1}\left(\frac{d}{pc}\right)^{2}}
\end{equation}
 where $\mathrm{S_{\nu}}$ is the integrated flux density, $\mathrm{\nu}$ is the frequency of the observation, $\mathrm{T_{e}}$ is the electron temperature, and d is the distance to the source. This was converted to the number of O-type stars assuming that a typical O-type star (O7.5V) produces $\mathrm{N_{Lyc}=10^{49}\:s^{-1}}$ \citep{1996ApJ...460..914V}. Table \ref{tab:sscprops} shows the associated ionizing flux for each clump that has 3.6 cm emissions detected above the 3$\sigma$ ($\mathrm{0.11\:mJy/beam}$) threshold and Figure \ref{fig:sscforming} shows the  3 and 5 $\sigma$ contours \textcolor{black}{overlaid on} the CO emission data.

Clump 9 is of particular note as the only potential SSC-forming clump detected above the 5$\sigma$ ($\mathrm{.19\:mJy/beam}$ or 60 O-star) threshold. Figure \ref{fig:sscforming} shows that clump 9 resides within a larger area of thermal radio emission contoured at the 5$\mathrm{\sigma}$ level at the bottom right of the map. This detection of ionizing flux is in line with \citet{2022ApJ...928...57H} which found a $\mathrm{10^{5.8}\:M_{\odot}}$ embedded young massive cluster in the same area enclosed by clump 9. The embedded cluster also has associated $\mathrm{Pa_{\beta}}$, I band, and optical detections \citep{2010AJ....140...75W,2014ApJ...795..156W,2022ApJ...928...57H}, indicating that it is a more evolved area.

\textcolor{black}{While the presence of 3.6 cm emission in certain areas of the Antennae overlap suggests that some of the molecular clouds have massive stars emitting ionizing flux, the other areas are not necessarily quiescent. Studies of infrared \citep{2004ApJS..154..193W} and $\mathrm{H_{\alpha}}$ \citep{2014MNRAS.445.1412Z} emissions show high SFRs in all of the SGMCs in the Antennae overlap. However, the lack of 3.6 cm emission in many of the CO-bright molecular clouds implies that the star formation occurring in those clouds has not affected the physical state of the ionized gas as much.}

\section{Comparative Analysis}\label{sec:companalasys}
SSCs are typically formed in galaxies that are experiencing a period of intense star formation called a starburst. Starbursting is typically caused by an interaction or merger with another galaxy \textcolor{black}{\citep{1991ApJ...370L..65B}}. These galaxies typically have star formation rate surface densities ($\mathrm{\Sigma_{SFR}}$) between 0.1 and 10 $\mathrm{M_{\odot}yr^{-1}kpc^{-2}}$. This is at least an order of magnitude higher than $\mathrm{\Sigma_{SFR}}$ in non-starbursting galaxies such as the Milky Way, which averages between $10^{-3}$ and $\mathrm{10^{-2}M_{\odot}yr^{-1}kpc^{-2}}$ \citep{2000ESASP.445...37L}. 

Although having an abundant reservoir of dense gas under high pressure has been connected to SSC formation, it does not guarantee that the galaxy has the capability to form SSCs. The central molecular zone (CMZ) of the Milky Way is also a high-pressure environment that contains a significant amount of dense gas \citep{2001ApJ...562..348O,quiescent_CMZ}, as well as a few young and massive star clusters such as the Arches and Quintuplet clusters. However, none of these young clusters meet the threshold for our definition of an SSC \citep{2010MNRAS.409..628H,2019ApJ...877...37R}. We seek to contextualize the Antennae overlap and identify key factors of SSC formation by comparing the Antennae overlap properties to those in a variety of environments. 

We compared the Antennae GMCs to those from 9 different environments observed in $\mathrm{^{12}CO}$. To minimize the effects of different resolutions between the different data sets, we limited our sample to data sets with spatial resolutions between 1-27 pc and spectral resolutions between 1-10 km/s in addition to deconvolving all radius and linewidth measurements (see Equations \eqref{eq:deconv} \& \eqref{eq:veldeconv}). We further discuss resolution effects in more depth in Appendix \ref{sec:res}.

The 9 data sets are separated into three separate groups for analysis: normal (N), starbursts (S), and galactic centers (GC). The N group consists of GMCs from areas with a normal amount of star formation and serves as a baseline to compare the other data sets to, group S is comprised of both dwarf and massive starbursting galaxies, and group GC consists of GMCs, in both starbursting and non-starbursting galaxies, that are located in the center of their host galaxies. The data sets for each of the environments and their groups are shown in Table \ref{tab:galaxies} along with their observed emission lines and resolutions.

In order to more directly compare the different GMCs, we re-calculated any physical properties that did not match our definitions described in Section \ref{sec:props}. This included defining the radius R and velocity dispersion $\mathrm{\sigma_{v}}$ with Equations \eqref{eq:deconv} and \eqref{eq:veldeconv}. We also only used masses calculated with an $\mathrm{CO\text{-}to\text{-}H_{2}}$ conversion factor. We further limited the mass calculations by only using the standard Galactic and starburst values of $\mathrm{X_{CO}=2\times10^{20}\;cm^{-2}(K\:km\:s^{-1})^{-1}}$ and $\mathrm{X_{CO}=0.5\times10^{20}\;cm^{-2}(K\:km\:s^{-1})^{-1}}$ respectively. For data sets using the CO(2-1) emission lines, values of $\mathrm{R_{CO(2-1)}}$ were taken from their original papers to convert to $\mathrm{L_{CO(1-0)}}$ and range from 0.65 to 0.93.

Of the data sets we used for the comparison environments, 8 of them used GMC identification and property calculation methods that were similar to our own. This allowed us to easily standardize the existing GMC catalogs to reflect the calculated property definitions for the molecular clouds as described above.  
\textcolor{black}{
\textcolor{black}{An} important comparison galaxy \textcolor{black}{in the nearby universe} is Henize 2-10, as another starburst galaxy that has had \textcolor{black}{numerous} SSCs detected in both the optical and radio wavelengths (Johnson \& Conti 2000; Johnson \& Kobulnicky 2003).  Given this similarity, we chose to re-define and compute cloud properties using identical methodology to the Antennae overlap.  We were unable to reproduce values similar to those published in  Imara \& Faesi (2019), and in particular the values in that paper have much smaller linewidths, masses, and effective radii, suggesting that they have decomposed into smaller sub-clouds than in our and other work (this is supported by their Figure~10 showing most of their clouds only subtending a small fraction of the velocity range along the line of sight).
Re-identifying and calculating properties for Henize 2-10 GMCs in this manner \textcolor{black}{yields} similar cloud properties to those found by \citet{2018ApJ...867..165B} using only the CPROPS algorithm on CO(3-2) observations of the same environment.}




\begin{deluxetable*}{ccccccccccc} 
\tablecaption{Data Sets Used in This Analysis\label{tab:galaxies}}
\tabletypesize{\scriptsize}
\tablehead{\colhead{Galaxy} & \colhead{CO Lines\tablenotemark{a}} & \colhead{Distance} & \colhead{Resolution} & \colhead{Channel} & \colhead{Galaxy Type} & \colhead{Region} & \colhead{Group} &\colhead{$\mathrm{X_{CO}}$} &\colhead{\textcolor{black}{Sensitivity}}&\colhead{Source}\\ 
\colhead{} & \colhead{} & \colhead{} & \colhead{pc} & \colhead{$\mathrm{km\:s^{-1}}$} & \colhead{} & \colhead{}&\colhead{} & \colhead{$\mathrm{10^{20}cm^{-2}(K\:km\:s^{-1})^{-1}}$}&\colhead{K} &\colhead{}} 

\startdata
Milky Way & 1-0 & 2-14 kpc & 0.5-3 & 1 & Barred Spiral  &Disk & N &2&0.09&1\\
M31 & 1-0 & 770 kpc & 25 & 1.3 & Barred Spiral & Spiral Arm & N &2&0.06&2\\
NGC 5068 & 2-1 & 5.2 Mpc & 26 & 2.5 & Barred Spiral & ... & N &2&0.21& 3 \\
Henize 2-10 & 1-0 & 8.7 Mpc & 22 & 1\tablenotemark{b} & BCD Starburst & ... & S &0.5&0.16& 4 \\
IC 10 & 1-0 & 950 kpc & 14-20 & 1,3,4 & Dwarf Starburst & Disk& S &0.5&0.17& 5 \\
NGC 5253& 2-1 & 3.15 Mpc & 3 & 1.47 & BCD Starburst & ... & S &2&0.13& 6 \\
NGC 4826& 2-1 & 2.2 Mpc & 27 & 2.5 & Spiral Starburst & ... & S &0.5&0.08& 3 \\
Antennae&2-1&22 Mpc&18&5&Merging Spirals&Overlap&S&0.5&1.7&This Work\\
NGC 4526 & 2-1 & 16.4 Mpc & 20 & 10 & Lenticular &Center&GC&2&0.71& 7 \\
Milky Way & 1-0 & 8.5 kpc & 1.4 & 2 & Barred Spiral  & Center&GC&0.5&0.42&8\\
\enddata
\tablenotetext{a}{All environments' emission lines are for $\mathrm{^{12}CO}$ with the exception of the Antennae which also uses $\mathrm{^{13}CO}$ data.}
\tablenotetext{b}{Later smoothed to 5 km/s}
\tablecomments{BCD (Blue Compact Dwarf)}.
\tablerefs{(1) \citet{article2}; (2)  \citet{Rosolowsky_2007}; (3) \citet{10.1093/mnras/stab085}; (4) \citet{2019ApJ...876..141I}; (5) \citet{2006ApJ...643..825L}; (6) \citet{Miura_2018}; (7) \citet{2015ApJ...803...16U}; (8) \citet{2001ApJ...562..348O}}
\end{deluxetable*}


\subsection{Size-Linewidth Relations}\label{sec:size-linewidth}
The relationship between a cloud's effective radius and linewidth is expected to follow a power law relationship of the form

\begin{equation}\label{eq:slw}
    \sigma_{v}=aR^{b}
\end{equation}
\citep{1981MNRAS.194..809L, 1987ApJ...319..730S}. 

\begin{figure*}
\centering
\includegraphics[width=.9\textwidth]{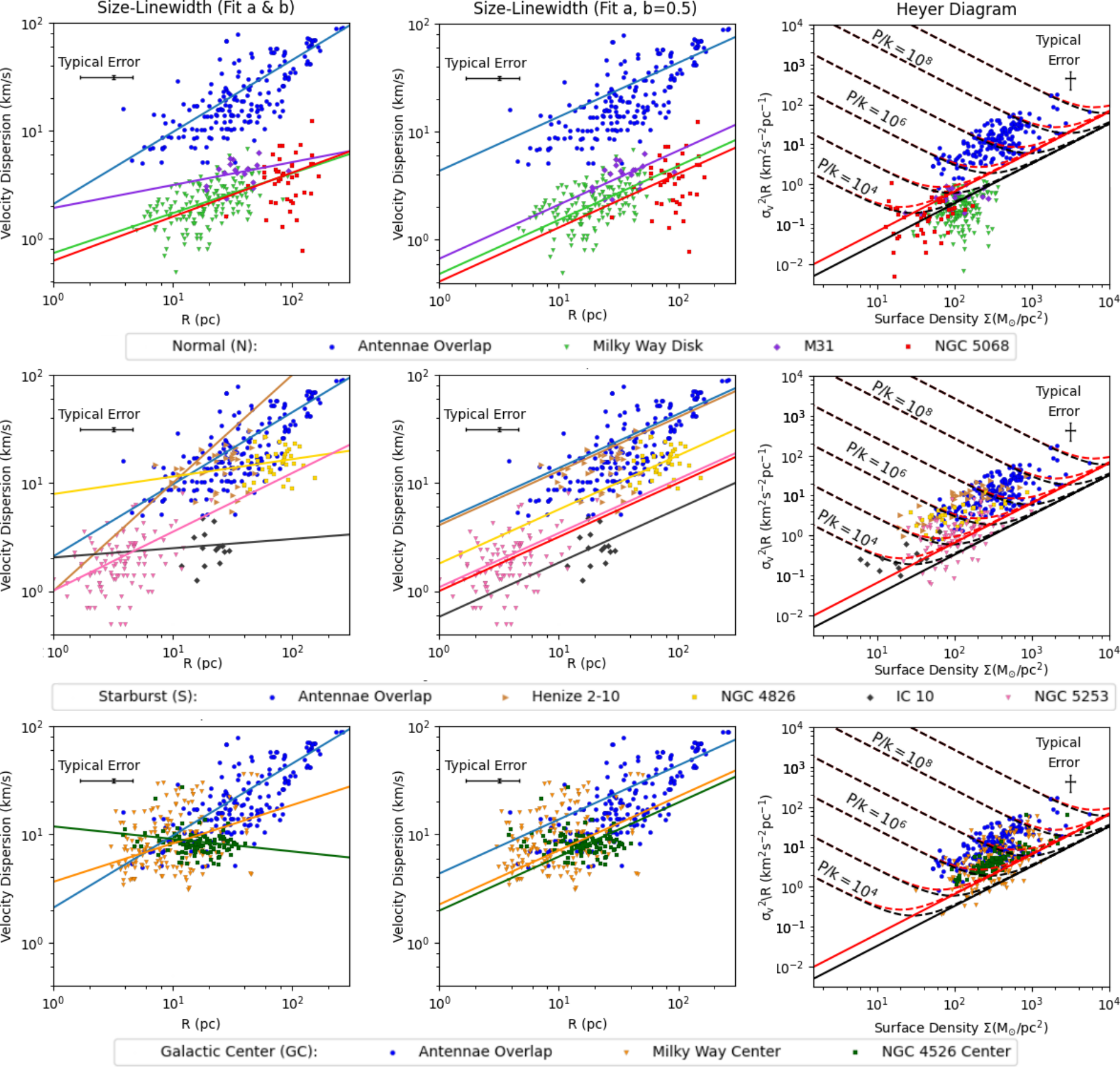}
\caption{Size-linewidth plots comparing the dendrogram structures detected in the $\mathrm{^{12}CO(2-1)}$ Antennae emission data to GMCs from galaxies in each group N, S, and GC. \emph{Left:} Size-linewidth plots where both the a and b parameters in Equation \eqref{eq:slw} were fit using the {\tt\string scipy\text{.}optimize\text{.}curve\_fit} implementation of the non-linear least squares method. \emph{Middle:} Size-linewidth plots where b is held constant at 0.5 and a is fit to the data. All of the fitted parameters are shown in Table \ref{tab:slw} \emph{Right:} Heyer relation plotting surface density $\Sigma$ against the velocity dispersions at given size scales $\mathrm{\sigma_{v}^{2}/R}$. The black lines represent virial equilibrium and the red lines represent free fall. The dashed black and red lines represent those same conditions accounting for external pressure ranging from $10^{4}-10^{9}$ K $\mathrm{cm^{-3}}$.}.
\label{fig:slw1}
\end{figure*}

Using Equation \eqref{eq:slw}, we fit lines to each of the emission lines we observed, as well as the emissions for the clouds, in Table \ref{tab:galaxies}. These are plotted in the leftmost column of Figure \ref{fig:slw1} by group with our $\mathrm{^{12}CO(2-1)}$ line in each plot representing the Antennae for comparison.

\begin{deluxetable}{cccccc}
\tablecaption{Size-Linewidth Line Fit Parameters\label{tab:slw}}
\tablehead{\colhead{Galaxy} & \colhead{Group} & \multicolumn{2}{c}{Fit a \& b} & \colhead{b=0.5}\\
\cline{3-4}
\colhead{} & \colhead{} & \colhead{(a)} & \colhead{(b)}  & \colhead{(a)}} 
\startdata
Milky Way Disk&N&0.74{\tiny$\pm0.01$}&0.37&0.48\\
M31 & N & 1.93{\tiny$\pm0.24$} & 0.21{\tiny$\pm0.03$} & 0.67\\
NGC 5068 & N & 0.62{\tiny$\pm0.11$} & 0.41{\tiny$\pm0.04$} & 0.41\\
Henize 2-10 & S & 2.86{\tiny$\pm0.01$}& 0.49{\tiny$\pm0.01$} & 4.07\\
IC 10 & S & 2.05{\tiny$\pm0.70$} & 0.09{\tiny$\pm0.11$} & 0.58\\
NGC 5253 & S & 1.01{\tiny$\pm0.01$} & 0.54{\tiny$\pm0.01$} & 1.08\\
NGC 4826 & S & 7.90{\tiny$\pm0.86$} & 0.16{\tiny$\pm0.03$} & 1.79\\
Antennae Clumps & S & 2.98{\tiny$\pm0.15$} & 0.55{\tiny$\pm0.01$} & 3.52{\tiny$\pm0.05$}\\
Antennae Dendro& S & 2.09{\tiny$\pm0.03$} & 0.67 & 4.34\\
NGC 4526 &GC& 11.84{\tiny$\pm0.20$} & -0.12{\tiny$\pm0.01$} & 1.97\\
Milky Way Center&GC& 3.64{\tiny$\pm0.06$} & 0.36{\tiny$\pm0.01$} & 2.25\\
\enddata
\tablecomments{Error is the standard error of the mean calculated from the covariance matrix. Errors less than 0.005 are not shown in the table above.}
\end{deluxetable}

Figure \ref{fig:slw1} shows the GMCs for the Antennae overlap having higher velocity dispersions for their sizes with their fitted lines having much larger slopes and intercepts than any other galaxy. Even using clumps, which Table \ref{tab:slw} shows having slightly smaller linewidths per given size scale than the dendrogram-identified structures, have larger linewidths than almost all of the other environments.

Among the three groups, the starburst and galactic center environments generally tend to have larger linewidths per size scale than normal environments, although the starburst galaxies have a much larger spread of linewidths than either of the other two groups.

Because linewidths tend to scale with surface density \citep{article2}, the effects of these enhanced linewidths per size scale in starbursts and galactic centers are better shown in relation to their surface densities. This is encapsulated by the revised size-linewidth relation set forth in \citet{article2} henceforth referred to as the Heyer relation
\begin{equation}\label{eq:heyer}
    \frac{\sigma_{v}^{2}}{R} \propto \Sigma.
\end{equation}

The rightmost column of Figure \ref{fig:slw1} plots this Heyer relation with solid black and red lines representing virial equilibrium and free-fall respectively. While the normal galaxies tend to fall around or below the virial equilibrium and free-fall lines, all of the starbursts and galactic centers tend to fall above those lines. This indicates that the GMCs in those galaxies are not entirely bound by gravity and must experience external pressure if they are bound as indicated by the dotted isobars. 

The effects of different spatial and spectral resolutions do also contribute small differences in the size-linewidth relations. In our investigation of resolution effects in Appendix \ref{sec:res}, we found that the fitted intercepts were only different by a factor of two or less. While this might change the order of similar environments, such as the Milky Way and NGC 5068, the overall trends tend to be larger and more on the scale of a magnitude. Additionally, the range of spatial and spectral resolutions spans both above and below the resolutions for the Antennae data. For example, IC 10 and NGC 4526 have similar spatial resolutions to the Antennae data, as well as each other, while their spectral resolutions are smaller and larger respectively. Despite this difference in spectra resolution, both of the galaxies have smaller linewidths per size scale than the Antennae overlap. This gives more credibility to the idea that these different size-linewidth relations are physically different, rather than simply due to the observation setups.

The large range of linewidths per size scales among the starburst galaxies is re-contextualized with the addition of surface density dependence. Each of the starburst galaxies are generally the same distance above the virial equilibrium and free-fall lines, with the exception of NGC 5253 which falls a bit lower and has some GMCs below the free-fall line. The larger physical result of the difference in velocity dispersions is the external pressures required for the GMCs to remain bound. These pressures span five orders of magnitude with pressures ranging between the $\mathrm{10^{4}}$ and $\mathrm{10^{9}}$ K $\mathrm{cm^{-3}}$ isobar lines. The extent that these GMCS differ from virial equilibrium and the external pressures required for them to remain bound is shown more directly in the pressure and virial parameter distributions in Section \ref{sec:propdist}.

\subsection{Property Distributions\label{sec:propdist}}
\begin{figure*}
\centering
\includegraphics[width=1.01\textwidth]{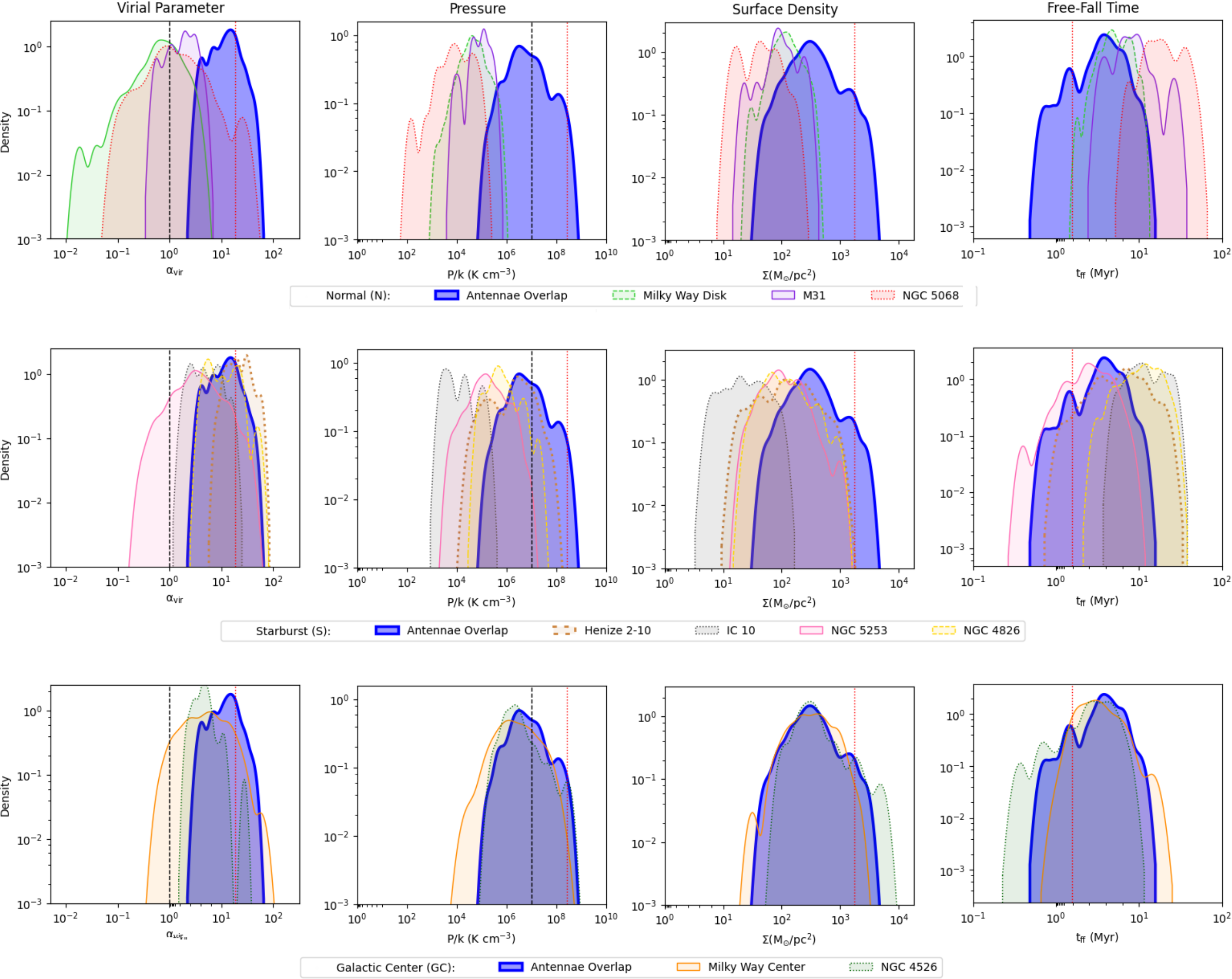}
\caption{Distributions of virial parameters, pressures, surface densities, and free-fall times (from left to right) for each of the GMC data sets represented as KDEs with smoothing parameters of
0.25 dex. The plots are split up in rows by environment type with the normal, starburst, and galactic center environments in the top, middle, and bottom rows respectively. The Antennae distributions are shown in each plot with the dotted red line representing where the Firecracker falls for each property. The pressure and virial parameter distributions also have additional dotted black lines that represent $\mathrm{P/k}$ = $\mathrm{10^{7}}$ K $\mathrm{cm^{-3}}$ and $\mathrm{\alpha_{vir}}$ = 1 respectively.}
\label{fig:alldist}
\end{figure*}

We compared the distributions of pressure, surface densities, free-fall times, and virial parameters of GMCs in the different galaxies to the Antennae overlap. These distributions are represented in Figure \ref{fig:alldist} as kernel density estimates (KDEs) using bandwidths of 0.25 dex.

Like the Heyer diagram in Figure \ref{fig:slw1}, Figure \ref{fig:alldist} shows that the Antennae GMCs have virial parameters that are much larger than virial equilibrium which has $\mathrm{\alpha_{vir}}$ = 1. \textcolor{black}{Because $\mathrm{\alpha_{vir}}$ depends on both R and $\mathrm{\sigma_{v}}$, and therefore is affected by observational biases, it is useful to view the virial parameters more generally.}

When calculating $\mathrm{\alpha_{vir}}$ directly, the distributions can be broadly split up into three different categories: mostly subvirial ($\mathrm{\overline{\alpha}_{vir}<<1}$), mostly supervirial ($\mathrm{\overline{\alpha}_{vir}>>1}$), generally virialized ($\mathrm{\overline{\alpha}_{vir}\sim1}$). Table \ref{tab:gmcvir} shows $\mathrm{\overline{\alpha}_{vir}}$ for each environment and their virialization category. The virialization categories were determined both by $\mathrm{\overline{\alpha}_{vir}}$ and as well as ensuring that they align with what is seen in Figures \ref{fig:slw1} and \ref{fig:alldist}.

\begin{deluxetable}{ccccc}
\tablecaption{GMC Virialization\label{tab:gmcvir}}
\tablehead{\colhead{Galaxy} & \colhead{Group} & \colhead{$\overline{\alpha}_{vir}$} & \colhead{1$\sigma$ Standard} & \colhead{Virialization} \\ &&&\colhead{Deviation}&} 
\startdata
Milky Way Disk & N & $0.92\pm0.05$ &0.69& virialized \\
M31 & N & $2.00\pm0.24$ &1.04& virialized \\
NGC 5068 & N & $2.32\pm0.59$ &3.94& virialized \\
Henize 2-10 & S & $24.50\pm2.07$ &12.42& supervirial \\
IC 10 & S & $5.92\pm1.00$ &3.72& supervirial \\
NGC 5253 & S & $5.14\pm0.55$ &5.54& supervirial \\
NGC 4826 & S & $12.88\pm1.51$ &9.18& supervirial \\
Antennae Overlap & S & $13.77\pm0.37$ &7.78& supervirial \\
Milky Way Center &GC& $7.22\pm0.56$ &7.23& supervirial \\
NGC 4526 Center&GC& $4.99\pm0.30$ &3.07& supervirial \\
\enddata
\end{deluxetable}

With these definitions, none of the GMC distributions are overall subvirial, although the Milky Way disk does have a large amount of GMCs with $\mathrm{\alpha_{vir}<1}$. While all of the distributions have some GMCs with $\mathrm{\alpha_{vir}}>1$, the starbursts and galactic centers are completely dominated by supervirial GMCs with the highest virial parameters. Among these supervirial environments is the Antennae overlap, which has the second highest $\mathrm{\overline{\alpha}_{vir}}$ after Henize 2-10.

With most of the $\mathrm{\alpha_{vir}}$ distributions falling around or above $\mathrm{\alpha_{vir}=1}$, the presence of external pressures is necessary for the GMCs to remain bound, \textcolor{black}{as well as to} collapse and form stars. These pressures are shown in the second column of Figure \ref{fig:alldist}. As previously seen in the Heyer diagram of Figure \ref{fig:slw1}, the Antennae overlap is among the highest-pressure environments along the galactic centers and Henize 2-10. 

\textcolor{black}{A K-S test suggests that \textcolor{black}{the distributions of the virial parameters in the Antennae overlap, Henize 2-10 and the centers of the Milky Way and NGC 4526} are very similar, with K-S statistics ranging between 0.13 and 0.31. The distributions of viral parameters for other galaxies tend to be less similar to the Antennae overlap with} K-S statistics $\sim 1$ and p-values $<<0.01$. A full table of the K-S test results is found in Appendix \ref{sec:kstest}.

\textcolor{black}{Pressure, surface density, and free fall time distributions also follow the same trend as the virial parameters. The GMCs at the centers of galaxies are the most similar to the Antennae overlap, while the normal galaxies are the least similar for all of the different properties. The starburst galaxies, however, are more ambiguous. Some of the starbursts, like Henize 2-10, are very similar to the Antennae overlap, while others are much more similar to normal galaxies.}

\textcolor{black}{This could be indicative of the wide range of environments that can form SSCs, though any conclusions based on this category are heavily dependent on the assumption that these properties are not significantly affected by the resolution of the observations and that our chosen $\mathrm{X_{CO}}$ and $\mathrm{R_{21}}$ values hold true.}

\textcolor{black}{While Appendix \ref{sec:res} shows that the observational resolution effects are small enough to not affect order-of-magnitude trends, they become much more important in less clear comparisons. The differences between the starburst galaxies, both to the other categories of galaxies and to each other, are small enough that the breakdown of the previously mentioned assumptions could give different results than what is shown in Figure \ref{fig:alldist}.}

\textcolor{black}{Our GMC property distributions are also affected by our choice of $\mathrm{X_{CO}}$ factors. Although we have chosen bimodal $\mathrm{X_{CO}}$ factors representing standard starbursts and non-starbursts, the $\mathrm{X_{CO}}$ is also known to vary with metallicity as well as with star formation activity \citep{2013ARA&A..51..207B}. This mainly affects the low-metallicity starbursts, Henize 2-10, IC 10, and NGC 5253, that have metallicity gradients that range from $\mathrm{\sim 0.25 Z_{\odot}\: to \: \sim Z_{\odot}}$ \citep{2006ApJ...643..825L, Miura_2018,2019ApJ...876..141I} and would bring the $\mathrm{X_{CO}}$ up to the standard \textcolor{black}{G}alactic $\mathrm{X_{CO}}$ value at the highest by Equation 18 in \citet{2013ARA&A..51..207B}. Increasing the estimated masses by up to a factor of four would not change our conclusions, as Henize 2-10 GMCs would only become more similar to the Antennae GMCs, and IC 10 and NGC 5253 would still be somewhere in the middle of all the different environments.}



\section{Discussion}\label{sec:discuss}
In addition to comparing the Antennae overlap to individual galaxies (Section \ref{sec:companalasys}), it is also useful to more broadly talk about how each of the environment types compare to the Antennae. Table \ref{tab:propertysummary} assigns each environment type with a similarity rating for each of the properties analyzed previously. These similarity ratings are calculated using the percent overlap between the \textcolor{black}{molecular cloud property distributions for each galaxy. We then rank the different groups of galaxies into ratings of} ``Very Dissimilar", ``Dissimilar", ``Similar", or ``Very Similar" \textcolor{black}{based on the average percent overlap of the galaxies in each environment type group.} These categories represent percent overlap in 25\% portions with the ``Very Dissimilar" being assigned to percent overlaps $\mathrm{<25\%}$ and the following labels corresponding to $\mathrm{<50\%}$, $\mathrm{<75\%}$, $\mathrm{<100\%}$ respectively.

\begin{deluxetable*}{c|cccccc|c}
\tablecaption{Properties by Environment Type\label{tab:propertysummary}} 
\tablehead{\colhead{\textbf{Group}} & \colhead{Size-Linewidths} & \colhead{$\mathrm{\sigma_{v}^{2}/R\:vs.\:\Sigma}$} & \colhead{$\mathrm{P_{e}}$} & \colhead{$\mathrm{\alpha_{vir}}$} & \colhead{$\mathrm{\Sigma}$} & \colhead{$\mathrm{t_{ff}}$}&\colhead{\textbf{Overall}}} 
\startdata
\textbf{Normal (N)} & \cellcolor{vds_color}Very Dissimilar & \cellcolor{vds_color}Very Dissimilar & \cellcolor{vds_color}Very Dissimilar & \cellcolor{vds_color} Very Dissimilar & \cellcolor{vds_color}Very Dissimilar & \cellcolor{ds_color}Dissimilar&\cellcolor{vds_color}\textbf{Very Dissimilar}\\
&&&&&&&\\
\textbf{Starbursts (S)}&\cellcolor{s_color}Similar&Very Similar & \cellcolor{s_color}Similar &Very Similar & \cellcolor{s_color}Similar & \cellcolor{s_color}Similar&\cellcolor{s_color}\textbf{Similar}\\
&&&&&&&\\
\textbf{Galactic Centers (GC)}& Very Similar & \cellcolor{ds_color}Dissimilar & Very Similar & \cellcolor{ds_color}Dissimilar & Very Similar & Very Similar&\textbf{Very Similar}\\
\enddata
\tablecomments{The level of similarity to the Antennae overlap was determined by calculating the percent overlap of the distributions. The average percent overlap for each property was calculated for all groups. The ``Very Dissimilar" label is assigned to properties with average percent overlap $\mathrm{<25\%}$ with the subsequent labels corresponding to $\mathrm{\leq50\%}$, $\mathrm{\leq75\%}$, and $\mathrm{\leq 100\%}$.}
\end{deluxetable*}

\textcolor{black}{Among these distinctions, it is important to remember that all of the properties in Table \ref{tab:propertysummary} depend on R and $\mathrm{\sigma_{v}}$ in some regard and therefore are impacted by resolution effects.} Although a full comparison of environments would require more homogeneous data, the existing datasets suggest that more extreme environments, such as galactic centers and some starbursts, are more similar to the Antennae overlap and the more normal galaxies are less similar.


\textcolor{black}{However, physical properties of the molecular clouds measured in this paper are not entirely indicative of the galaxies' SFRs. While the properties of Antennae overlap and Henize 2-10 are more similar to the centers of galaxies, several other starburst galaxies have properties that are much more similar to the more normal galaxies. For example, IC 10 appears to have very low-density molecular clouds, despite having a high star formation rate.} This could be due to differences in SFEs that, if true, would allow some of the galaxies to form SSCs despite having less abundant reservoirs of dense molecular gas. It is also important to note that Henize 2-10, which is the starburst with the most similar physical properties to the Antennae overlap of the galaxies in this analysis. In addition to having a large population of fully-emerged SSCs, it also has a range of evolutionary stages, including clusters still deeply embedded in their birth material. It is possible that instead of forming SSCs, some of the galaxies are instead forming lower-mass clusters, leading to high overall SFRs without necessarily forming high-mass clusters at the moment.

On the other end of galaxies with physical properties that do not match their SFR, galactic centers with the extreme physical properties seen in Figures \ref{fig:slw1} and \ref{fig:alldist} do not have proportionately high SFRs. \textcolor{black}{\citet{2004ApJS..154..193W} found a similar trend in both galactic centers of the Antennae galaxies with the SFR in the Antennae being highest in the outer arms and the overlap region, despite there being plenty of dense gas in the galactic centers.} This supports the idea that, although it is important to have enough dense, pressurized gas to form massive star clusters, it is not the only necessary condition \citep{2004ApJS..154..193W,2013MNRAS.429..987L}. One possible additional requirement is turbulence-induced shocks that promote the formation of star formation structures such as filaments \citep{2018PASJ...70S..53I}.

\section{Conclusion}\label{sec:conc}
We present high-resolution ($\sim$ 0.1") ALMA observations of the overlap region of the Antennae galaxies (NGC 4038/39). The emission from $\mathrm{^{12}CO(2-1)}$, $\mathrm{^{12}CO(3-2)}$, and $\mathrm{^{13}CO(2-1)}$ were used to characterize GMCs and their environments which we then compared to GMCs from other galaxies summarized in Table \ref{tab:galaxies}. In this paper we:

\begin{itemize}
  \item Decomposed the emission into singular, non-overlapping clumps using {\tt\string quickclump} \citep{2017ascl.soft04006S} as well as hierarchical structures using {\tt\string astrodendro} \citep{2008ApJ...679.1272M}. This yielded 72 clumps and 206 dendrogram structures which consisted of 124 leaves, 72 branches, and 10 trunks. (Section \ref{sec:structure})
  \item Calculated the velocity dispersion, effective radius, mass, surface density, pressure, and $\mathrm{\alpha_{vir}}$ of each GMC in the Antennae overlap region summarized in Tables \ref{tab:12ClumpProps} - \ref{tab:13DendroProps}. (Section \ref{sec:props} and Appendix \ref{sec:catalog})
  \item Compared the physical properties Antennae overlap GMCs to those from a variety of environments which are identified in Table \ref{tab:galaxies}.
\end{itemize}

Throughout our comparative analysis (Sections \ref{sec:companalasys} and \ref{sec:discuss}), we found several trends between the type of environment and its GMC properties. Our primary findings are as follows:
\begin{itemize}
  \item The GMCs in the Antennae overlap have average densities, pressures, and virial parameters of $\mathrm{\overline{\Sigma}}=440.64\:M_{\odot}pc^{-2}$, $\mathrm{\overline{P_{e}}/k=2.11\times 10^{7}\:K\:cm^{-3}}$, and $\mathrm{\overline{\alpha}_{vir}=13.77}$. Six of the clumps consistently lie outside of the normal physical property distributions, one of which is the pre-SSC-forming cloud called the Firecracker. These clumps have particularly high surface densities and pressures. They also meet the mass, radius, and pressure for an SSC progenitor cloud along with eleven other clumps. (Section \ref{sec:antprops})
  \item Comparing the size-linewidth relations of the Antennae overlap to those for normal GMCs (N), starburst galaxies (S), and galactic centers (GC), the Antennae overlap has much higher fitted intercepts in a size-linewidth plot than most other environments with the exception of Henize 2-10, another SSC-forming starburst galaxy. All of the galaxies in group N and some in group S had the lowest kinetic energy while GMCs in the GC group had some of the highest kinetic energies. Group S is split with some of the starbursts having high kinetic energies like the Antennae overlap and some having lower energies closer to those of normal galaxies.(Section \ref{sec:size-linewidth})
  \item \textcolor{black}{Although a full comparison of environments would require more homogeneous data, the Antennae overlap and the galactic centers tend to have higher virial parameters, surface densities, and pressures than the normal galaxies. The other starburst galaxies tend to fall between the galactic centers and normal galaxies. (Sections \ref{sec:propdist} and \ref{sec:discuss})}

\end{itemize}

\begin{acknowledgments}
    This paper makes use of the following ALMA data: ADS/JAO.ALMA\#2015.1.00977.S and ADS/JAO.ALMA\#2016.1.00924.S. ALMA is a partnership of ESO (representing its member states), NSF (USA) and NINS (Japan), together with NRC (Canada), MOST and ASIAA (Taiwan), and KASI (Republic of Korea), in cooperation with the Republic of Chile. The Joint ALMA Observatory is operated by ESO, AUI/NRAO and NAOJ.The National Radio Astronomy Observatory is a facility of the National Science Foundation operated under cooperative agreement by Associated Universities, Inc.
\end{acknowledgments}

\facilities{ALMA - Atacama Large Millimeter Array}

\software{Astrodendro \citep{Rosolowsky_2008},
          Astropy \citep{2013A&A...558A..33A,2018AJ....156..123A},
          CASA          \citep{2007ASPC..376..127M},
          Matplotlib \citep{Hunter:2007},
          NumPy \citep{harris2020array}, Pandas \citep{mckinney-proc-scipy-2010},
          Papanda (\url{https://github.com/Ruslan-Nazarov/papanda.git})
          Quickclump \citep{2017ascl.soft04006S},
          Seaborn \citep{Waskom2021},
          SciPy \citep{2020NatMe..17..261V}
          }

\appendix
\section{GMC Catalogs}\label{sec:catalog}
GMC catalogs for each emission line observed for the Antennae overlap. Each emission has two catalogs: one using clump decomposition and the other using dendrogram decomposition. Unless otherwise stated, the properties are calculated and labeled as described in Section \ref{sec:props}.

\begin{deluxetable}{ccccccccccccc}[h]
\tablecaption{$\mathrm{^{12}CO(2-1)}$ Clump Properties\label{tab:12ClumpProps}}
\tablenum{A1}
\tablehead{\colhead{ncl} & \colhead{Type} & \colhead{log $\mathrm{L_{CO}}$} & \colhead{CO max} & \colhead{log $\mathrm{M_{lum}}$} & \colhead{log $\mathrm{M_{LTE}}$} & \colhead{$\mathrm{\sigma_{v}}$} & \colhead{ maj x min} & \colhead{R} & \colhead{log area} & \colhead{log $\mathrm{\Sigma}$} & \colhead{log $\mathrm{P_{e}/k}$} & \colhead{log $\mathrm{\alpha_{vir}}$} \\ 
\colhead{} & \colhead{} & \colhead{(K $\mathrm{km\:s^{-1}}$)} & \colhead{(K)} & \colhead{($\mathrm{M_{\odot}}$)} & \colhead{($\mathrm{M_{\odot}}$)} & \colhead{($\mathrm{km\:s^{-1}}$)} & \colhead{(pc x pc)} & \colhead{(pc)} & \colhead{($\mathrm{pc^{2}}$)} & \colhead{($\mathrm{M_{\odot}\:pc^{-2}}$)} & \colhead{($\mathrm{K cm^{-3}}$)} & \colhead{} }

\startdata
1 & 2 & 6.51$\pm$2.43 & 28.4 & 6.53$\pm$5.53 & 6.40& 1.67$\pm$0.17 & 15.4x22.7 & 24.8$\pm$2.45 & 3.72 & 3.24$\pm$2.24 & 8.45$\pm$7.95 & 18.6$\pm$2.87 \\
2 & 2 & 6.81$\pm$2.52 & 34.8 & 6.81$\pm$5.81 & 7.05 & 33.6$\pm$1.16 & 26.2x47.7 & 54.7$\pm$7.97 & 4.08 & 2.83$\pm$1.83 & 7.41$\pm$7.07 & 11.2$\pm$2.12 \\
3 & 2 & 6.87$\pm$2.65 & 21.1 & 6.89$\pm$5.89 & 6.94 & 45.6$\pm$1.99 & 66.5x42.8 & 84.9$\pm$16.4 & 4.37 & 2.53$\pm$1.53 & 7.19$\pm$6.96 & 26.6$\pm$6.22 \\
4 & 2 & 6.19$\pm$2.24 & 27.4 & 6.19$\pm$5.19 & 6.45 & 20.3$\pm$0.85 & 30.1x17.6 & 33.0$\pm$12.3 & 3.67 & 2.65$\pm$1.68 & 7.01$\pm$7.06 & 10.2$\pm$4.05 \\
5 & 2 & 6.74$\pm$2.6 & 19.6 & 6.76$\pm$5.76 & 6.8 & 34.1$\pm$1.47 & 43.9x32.4 & 58.7$\pm$16.4 & 4.3 & 2.72$\pm$1.73 & 7.29$\pm$7.21 & 13.83$\pm$4.26 \\
6 & 2 & 6.27$\pm$2.36 & 18.1 & 6.29$\pm$5.29 & 6.46 & 21.9$\pm$0.88 & 36.7x34.3 & 54.9$\pm$12.1 & 2.99 & 2.31$\pm$1.32 & 6.52$\pm$6.35 & 15.6$\pm$3.98 \\
7 & 2 & 6.27$\pm$2.21 & 34.7 & 6.28$\pm$5.28 & 6.73 & 16.9$\pm$0.69 & 29.0x18.7 & 33.5$\pm$10.8 & 3.68 & 2.73$\pm$1.75 & 6.93$\pm$6.92 & 5.79$\pm$2.01 \\
8 & 2 & 6.09$\pm$2.26 & 17.0 & 6.11$\pm$5.11 & 6.34 & 30.2$\pm$1.26 & 20.7x23.8 & 31.4$\pm$6.92 & 3.76 & 2.62$\pm$1.63 & 7.34$\pm$7.17 & 25.9$\pm$6.62 \\
9 & 2 & 5.7$\pm$1.99 & 24.9 & 5.71$\pm$4.71 & 6.06 & 15.3$\pm$0.67 & 9.54x14.8 & 7.89$\pm$12.4 & 3.48 & 3.42$\pm$3.21 & 8.16$\pm$8.84 & 4.14$\pm$6.52 \\
10 & 2 & 6.69$\pm$2.41 & 33.5& 6.7$\pm$5.7 & 7.06 & 24.9$\pm$0.95 & 39.0x23.8 & 46.2$\pm$16.0 & 4.00 & 2.87$\pm$1.88 & 7.26$\pm$7.28 & 6.62$\pm$2.43 \\
\enddata
\tablecomments{Table of physical properties of the $\mathrm{^{12}}$CO(2-1) emission of the Antennae overlap separated into structures by {\tt\string quickclump} where ``ncl" is the cloud number and ``Type" is the type of the cloud where 0 is a dendrogram trunk, 1 is a dendrogram branch, and 2 is a quickclump clump or dendrogram leaf. ``CO max" is the local maximum CO luminosity, and ``maj x min" are the major and minor axes of the ellipse fitted to the HWHM of the clump. All other properties are as defined in Section \ref{sec:props}.  Table A1 is published in its entirety in the machine-readable format. The first 10 (out of 72) clumps are shown here for guidance regarding its form and content.}
\end{deluxetable}

\begin{deluxetable}{ccccccccccccc}[h]
\tablecaption{$\mathrm{^{12}CO(2-1)}$ Dendrogram Properties\label{tab:12DendroProps}}
\tablenum{A2}
\tablehead{\colhead{ncl} & \colhead{Type} & \colhead{log $\mathrm{L_{CO}}$} & \colhead{CO max} & \colhead{log $\mathrm{M_{lum}}$} & \colhead{log $\mathrm{M_{LTE}}$} & \colhead{$\mathrm{\sigma_{v}}$} & \colhead{ maj x min} & \colhead{R} & \colhead{log area} & \colhead{log $\mathrm{\Sigma}$} & \colhead{log $\mathrm{P_{e}/k}$} & \colhead{log $\mathrm{\alpha_{vir}}$} \\ 
\colhead{} & \colhead{} & \colhead{(K $\mathrm{km\:s^{-1}}$)} & \colhead{(K)} & \colhead{($\mathrm{M_{\odot}}$)} & \colhead{($\mathrm{M_{\odot}}$)} & \colhead{($\mathrm{km\:s^{-1}}$)} & \colhead{(pc x pc)} & \colhead{(pc)} & \colhead{($\mathrm{pc^{2}}$)} & \colhead{($\mathrm{M_{\odot}\:pc^{-2}}$)} & \colhead{($\mathrm{K cm^{-3}}$)} & \colhead{} } 

\startdata
0 & 0 & 7.74$\pm$3.04 & 34.8 & 7.74$\pm$6.74 & 7.89 & 89.7$\pm$5.08 & 258x102 & 263$\pm$89.7 & 5.13 & 2.4$\pm$1.4 & 7.15$\pm$7.17 & 44.8$\pm$16.7 \\
1 & 1 & 7.58$\pm$2.95 & 34.8 & 7.58$\pm$6.58 & 7.77 & 88.4$\pm$4.84 & 243x85.9 & 234$\pm$90.7 & 4.91 & 2.34$\pm$1.34 & 7.13$\pm$7.2 & 55.7$\pm$23.1 \\
2 & 1 & 7.38$\pm$2.86 & 34.8 & 7.37$\pm$6.37 & 7.53 & 55.6$\pm$2.28 & 119x92.6 & 170$\pm$50.2 & 4.65 & 2.42$\pm$1.42 & 6.94$\pm$6.9 & 25.67$\pm$8.28 \\
3 & 1 & 7.62$\pm$2.97 & 34.8 & 7.62$\pm$6.62 & 7.79 & 88.9$\pm$4.91 & 259x96.1 & 256$\pm$151 & 4.93 & 2.31$\pm$1.31 & 7.06$\pm$7.31 & 56.3$\pm$34.3 \\
4 & 1 & 7.25$\pm$2.77 & 34.8 & 7.24$\pm$6.24 & 7.42 & 54.3$\pm$2.29 & 29.0x102 & 86.5$\pm$60.3 & 4.49 & 2.87$\pm$1.88 & 7.67$\pm$7.99 & 16.9$\pm$12 \\
5 & 1 & 7.21$\pm$2.74 & 34.8 & 7.2$\pm$6.2 & 7.39 & 51.9$\pm$2.07 & 28.4x102 & 85.6$\pm$58.5 & 4.42 & 2.84$\pm$1.85 & 7.6$\pm$7.92 & 16.8$\pm$11.7 \\
6 & 1 & 7.14$\pm$2.69 & 34.8 & 7.13$\pm$6.13 & 7.34 & 50.4$\pm$2.03 & 28.3x100 & 84.7$\pm$62.6 & 4.37 & 2.78$\pm$1.79 & 7.52$\pm$7.87 & 18.4$\pm$13.8 \\
7 & 2 & 6.46$\pm$2.35 & 24.8 & 6.45$\pm$5.45 & 6.30 & 32.0$\pm$1.29 & 17.5x33.2 & 34.9$\pm$13.5 & 3.70 & 2.87$\pm$1.89 & 7.6$\pm$7.67 & 14.64$\pm$5.97 \\
8 & 1 & 6.57$\pm$2.47 & 27.3 & 6.57$\pm$5.57 & 6.65 & 52.6$\pm$2.57 & 56.3x15.8 & 45.2$\pm$27.0 & 4.02 & 2.76$\pm$1.79 & 7.81$\pm$8.07 & 39.1$\pm$24.0 \\
9 & 0 & 7.76$\pm$3.02 & 28.4 & 7.78$\pm$6.78 & 7.86 & 124$\pm$5.83 & 19.7x30.7 & 35.8$\pm$6.91 & 5.28 & 4.17$\pm$3.18 & 1.06$\pm$9.84 & 10.6$\pm$2.5 \\
10 & 1 & 7.64$\pm$2.99 & 21.1 & 7.66$\pm$6.66 & 7.76 & 69.3$\pm$2.87 & 144x62.0 & 153$\pm$63.6 & 5.18 & 2.8$\pm$1.8 & 7.56$\pm$7.66 & 18.6$\pm$8.12 \\
\enddata
\tablecomments{Table of physical properties of the $\mathrm{^{12}}$CO(2-1) emission of the Antennae overlap separated into structures by {\tt\string astrodendro} where ``ncl" is the cloud number and ``Type" is the type of the cloud where 0 is a dendrogram trunk, 1 is a dendrogram branch, and 2 is a quickclump clump or dendrogram leaf. ``CO max" is the local maximum CO luminosity, and ``maj x min" are the major and minor axes of the ellipse fitted to the HWHM of the clump. All other properties are as defined in Section \ref{sec:props}. Table A2 is published in its entirety in the machine-readable format. The first 10 (out of 206) dendrogram structures are shown here for guidance regarding its form and content.}
\end{deluxetable}

\begin{deluxetable}{cccccccccccc}[h]
\tablecaption{$\mathrm{^{13}CO(2-1)}$ Clump Properties\label{tab:13ClumpProps}}
\tablenum{A3}
\tablehead{\colhead{ncl} & \colhead{Type} & \colhead{log $\mathrm{L_{CO}}$} & \colhead{CO max} & \colhead{log $\mathrm{M_{lum}}$}  & \colhead{$\mathrm{\sigma_{v}}$} & \colhead{ maj x min} & \colhead{R} & \colhead{log area} & \colhead{log $\mathrm{\Sigma}$} & \colhead{log $\mathrm{P_{e}/k}$} & \colhead{log $\mathrm{\alpha_{vir}}$} \\ 
\colhead{} & \colhead{} & \colhead{(K $\mathrm{km\:s^{-1}}$)} & \colhead{(K)} & \colhead{($\mathrm{M_{\odot}}$)} & \colhead{($\mathrm{km\:s^{-1}}$)} & \colhead{(pc x pc)} & \colhead{(pc)} & \colhead{($\mathrm{pc^{2}}$)} & \colhead{($\mathrm{M_{\odot}\:pc^{-2}}$)} & \colhead{($\mathrm{K cm^{-3}}$)} & \colhead{} }

\startdata
1 & 2 & 5.28$\pm$1.05 & 1.85 & 6.72$\pm$5.72 &  48.1$\pm$2.2 & 19.2x29.2 & 34.2$\pm$5.64 & 3.72 & 3.15$\pm$2.16 & 8.25$\pm$7.96 & 17.5$\pm$3.74 \\
2 & 2 & 5.87$\pm$1.69 & 8.45 & 7.33$\pm$6.33 & 29.2$\pm$1.08 & 17.8x35.7 & 37.02$\pm$8.42 & 4.08 & 3.69$\pm$2.7 & 8.32$\pm$8.16 & 1.73$\pm$0.45 \\
3 & 2 & 5.84$\pm$1.16 & 2.65 & 7.27$\pm$6.27 & 34.2$\pm$0.96 & 66.4x42.6 & 84.7$\pm$28.3 & 4.37 & 2.92$\pm$1.92 & 7.32$\pm$7.33 & 6.17$\pm$2.17 \\
4 & 2 & 5.3$\pm$1.46 & 5.10 & 6.76$\pm$5.76 & 18.4$\pm$0.82 & 17.7x18.3 & 23.3$\pm$4.13 & 3.67 & 3.53$\pm$2.54 & 7.95$\pm$7.69 & 1.59$\pm$0.35 \\
5 & 2 & 5.74$\pm$1.1 & 2.23 & 7.18$\pm$6.18 & 26.2$\pm$0.52 & 47.2x4 & 68.4$\pm$9.54 & 4.28 & 3.01$\pm$2.01 & 7.27$\pm$6.91 & 3.62$\pm$0.64 \\
6 & 2 & 5.36$\pm$1.18 & 2.62 & 6.79$\pm$5.79 & 21.2$\pm$0.9 & 35.7x34.3 & 54.1$\pm$12.3 & 2.99 & 2.83$\pm$1.83 & 7.02$\pm$6.86 & 4.56$\pm$1.19 \\
7 & 2 & 5.53$\pm$1.63 & 9.00 & 6.79$\pm$5.79 & 14.9$\pm$0.44 & 24x15.7 & 26.18$\pm$4.68 & 3.68 & 3.46$\pm$2.47 & 7.65$\pm$7.39 & 1.09$\pm$0.23 \\
8 & 2 & 5.23$\pm$1.32 & 3.7 & 6.66$\pm$5.66 & 30.3$\pm$1.82 & 21.7x16.0 & 24.7$\pm$5.58 & 3.76 & 3.38$\pm$2.4 & 8.21$\pm$8.06 & 5.7$\pm$1.56 \\
9 & 2 & 4.91$\pm$1.36 & 4.55 & 6.17$\pm$5.17 & 14.2$\pm$0.52 & 10.1x15.6 & 10.3$\pm$9.80 & 3.48 & 3.65$\pm$3.12 & 8.2$\pm$8.66 & 1.62$\pm$1.55 \\
10 & 2 & 5.85$\pm$1.78 & 8.30 & 7.11$\pm$6.11 & 19.1$\pm$0.39 & 39.9x18.8 & 40.9$\pm$15.9 & 4.00 & 3.39$\pm$2.41 & 7.61$\pm$7.68 & 1.32$\pm$0.53 \\
\enddata
\tablecomments{Table of physical properties of the $\mathrm{^{13}}$CO(2-1) emission of the Antennae overlap separated into structures by {\tt\string quickclump} where ``ncl" is the cloud number and ``Type" is the type of the cloud where 0 is a dendrogram trunk, 1 is a dendrogram branch, and 2 is a quickclump clump or dendrogram leaf. ``CO max" is the local maximum CO luminosity, and ``maj x min" are the major and minor axes of the ellipse fitted to the HWHM of the clump. All other properties are as defined in Section \ref{sec:props}. Table A3 is published in its entirety in the machine-readable format. The first 10 (out of 72) clumps are shown here for guidance regarding its form and content.}
\end{deluxetable}

\begin{deluxetable}{cccccccccccc}[h]
\tablecaption{$\mathrm{^{13}CO(2-1)}$ Dendrogram Properties\label{tab:13DendroProps}}
\tablenum{A4}
\tablehead{\colhead{ncl} & \colhead{Type} & \colhead{log $\mathrm{L_{CO}}$} & \colhead{CO max} & \colhead{log $\mathrm{M_{lum}}$}  & \colhead{$\mathrm{\sigma_{v}}$} & \colhead{ maj x min} & \colhead{R} & \colhead{log area} & \colhead{log $\mathrm{\Sigma}$} & \colhead{log $\mathrm{P_{e}/k}$} & \colhead{log $\mathrm{\alpha_{vir}}$} \\ 
\colhead{} & \colhead{} & \colhead{(K $\mathrm{km\:s^{-1}}$)} & \colhead{(K)} & \colhead{($\mathrm{M_{\odot}}$)} & \colhead{($\mathrm{km\:s^{-1}}$)} & \colhead{(pc x pc)} & \colhead{(pc)} & \colhead{($\mathrm{pc^{2}}$)} & \colhead{($\mathrm{M_{\odot}\:pc^{-2}}$)} & \colhead{($\mathrm{K cm^{-3}}$)} & \colhead{} }

\startdata
0 & 0 & 6.75$\pm$2.09 & 8.45 & 8.21$\pm$7.21 & 94.7$\pm$7.41 & 84.8x218 & 220$\pm$122 & 5.12 & 3.03$\pm$2.03 & 7.9$\pm$8.13 & 14.1$\pm$8.22 \\
1 & 1 & 6.63$\pm$2.06 & 8.45 & 8.08$\pm$7.08 & 92.3$\pm$7.24 & 68.3x216 & 197$\pm$149 & 4.91 & 3$\pm$2 & 7.9$\pm$8.26 & 16$\pm$12.45 \\
2 & 1 & 6.39$\pm$1.91 & 8.45 & 7.85$\pm$6.85 & 54.8$\pm$2.89 & 87.5x63.3 & 120$\pm$34.1 & 4.65 & 3.2$\pm$2.2 & 7.86$\pm$7.8 & 5.86$\pm$1.88 \\
3 & 1 & 6.65$\pm$2.06 & 8.45 & 8.11$\pm$7.11 & 90.8$\pm$6.94 & 78.1x207 & 206$\pm$53.2 & 4.92 & 2.98$\pm$1.98 & 7.85$\pm$7.75 & 15.4$\pm$4.86 \\
4 & 1 & 6.28$\pm$1.88 & 8.45 & 7.73$\pm$6.73 & 54.9$\pm$2.91 & 28.8x87.6 & 79.8$\pm$52.8 & 4.49 & 3.43$\pm$2.45 & 8.27$\pm$8.57 & 5.14$\pm$3.48 \\
5 & 1 & 6.24$\pm$1.88 & 8.45 & 7.7$\pm$6.7 & 53.2$\pm$2.91 & 29.2x87.2 & 80.1$\pm$54.8 & 4.42 & 3.39$\pm$2.41 & 8.21$\pm$8.52 & 5.28$\pm$3.69 \\
6 & 1 & 6.18$\pm$1.87 & 8.45 & 7.63$\pm$6.63 & 51.7$\pm$2.95 & 28.7x87.2 & 79.4$\pm$57.6 & 4.37 & 3.34$\pm$2.35 & 8.13$\pm$8.47 & 5.71$\pm$4.23 \\
7 & 2 & 5.16$\pm$1.15 & 2.84 & 6.62$\pm$5.62 & 24.9$\pm$1.23 & 18.3x37.4 & 38.7$\pm$14.6 & 3.70 & 2.95$\pm$1.97 & 7.42$\pm$7.47 & 6.62$\pm$2.66 \\
8 & 1 & 5.53$\pm$1.49 & 5.10 & 6.98$\pm$5.98 & 19.8$\pm$1.24 & 24.1x18.3 & 29.2$\pm$6.82 & 4.02 & 3.56$\pm$1.57 & 7.95$\pm$7.81 & 1.38$\pm$0.39 \\
9 & 0 & 6.77$\pm$1.54 & 3.71 & 8.2$\pm$7.2 & 132$\pm$9.15 & 155x126 & 226$\pm$43.8 & 5.28 & 3$\pm$0 & 8.15$\pm$7.94 & 28.5$\pm$7.36 \\
10 & 1 & 6.67$\pm$1.54 & 3.71 & 8.1$\pm$7.1 & 67.2$\pm$3.04 & 155x65.9 & 163$\pm$93.0 & 5.18 & 3.18$\pm$2.18 & 7.89$\pm$8.12 & 6.77$\pm$3.97 \\
\enddata
\tablecomments{Table of physical properties of the $\mathrm{^{13}}$CO(2-1) emission of the Antennae overlap separated into structures by {\tt\string astrodendro} where ``ncl" is the cloud number and ``Type" is the type of the cloud where 0 is a dendrogram trunk, 1 is a dendrogram branch, and 2 is a quickclump clump or dendrogram leaf. ``CO max" is the local maximum CO luminosity, and ``maj x min" are the major and minor axes of the ellipse fitted to the HWHM of the clump. All other properties are as defined in Section \ref{sec:props}.  Table A4 is published in its entirety in the machine-readable format. The first 10 (out of 72) clumps are shown here for guidance regarding its form and content.}
\end{deluxetable}

\section{K-S Tests}\label{sec:kstest}
In order to determine how similar the different physical property distributions in the various environments are to the Antennae overlap distribution, two-sample Kolmogorov–Smirnov (K-S) tests were performed on the GMC catalogs using {\tt\string scipy\text{.}stats\text{.}ks\_2samp}. The K-S test produces a K-S statistic that ranges from 0-1 and quantifies the distance between the two distributions, as well as a p-value to determine the likelihood that the data could be drawn from the same distribution. In addition to comparing each of the environments to the Antennae overlap, K-S tests were also performed between the GMCs from the NGC 4526 and Milky Way centers, NGC 4826 and NGC 5253, and Henize 2-10 and IC 10. The results of the K-S tests are shown in Table \ref{tab:ks} below.

\begin{deluxetable}{ccccccccccccccccc}[h]
\tablecaption{K-S Tests\label{tab:ks}}
\tablenum{B1}
\tablehead{\colhead{Galaxy 1} & \colhead{Galaxy 2} & \multicolumn{2}{c}{$\mathrm{\Sigma}$} && \multicolumn{2}{c}{$\mathrm{P_{e}/k}$} && \multicolumn{2}{c}{$\mathrm{\alpha_{vir}}$}&&\multicolumn{2}{c}{tff}\\ 
\cline{3-4}  \cline{6-7} \cline{9-10}  \cline{12-13}
\colhead{} & \colhead{} & \colhead{(K-S)} & \colhead{(p-value)} && \colhead{(K-S)} & \colhead{(p-value)} && \colhead{(K-S)} & \colhead{(p-value)} && \colhead{(K-S)} & \colhead{(p-value)}}

\startdata
Milky Way Disk (N) & Antennae (S) & 0.650 & 0.00 && 0.965 & 0.00 && 0.994 & 0.00 && 0.376 & 4.30E-6\\
M31 (N) & Antennae (S) & 0.678 & 6.45E-07 && 0.984 & 2.03E-25 && 0.947 & 2.67E-15 && 0.661 & 1.41E-06 \\
NGC 5068 (N) & Antennae (S) & 0.889 & 3.05E-25 && 1.00 & 0.00 && 0.847 & 1.34E-21 && 0.967 & 5.12E-39 \\
Henize 2-10 (S) & Antennae (S) & 0.425 & 2.78E-04 && 0.278 & 0.04 && 0.458 & 6.47E-05 && 0.454 & 7.91E-05 \\
IC 10 (S) & Antennae (S) & 0.951 & 2.43E-12 && 1.00 & 3.78E-15 && 0.617 & 1.25E-04 && 0.896 & 1.59E-10\\
NGC 5253 (S) & Antennae (S) & 0.588 & 6.88E-13 && 0.778 & 7.70E-25 && 0.633 & 5.55E-15 && 0.437 & 4.14E-7 \\
NGC 4826 (S) & Antennae (S) & 0.581 & 9.33E-08 && 0.658 & 5.62E-10 && 0.231 & \textbf{0.142} && 0.778 & 1.83E-14 \\
Milky Way Center (GC) & Antennae (S) & 0.092 & \textbf{0.803} && 0.291 & 7.62E-04 && 0.477 & 9.88E-10 && 0.320 & 1.54E-4 \\
NGC 4526 Center (GC) & Antennae (S) & 0.118 & \textbf{0.608} && 0.306 & 1.12E-03 && 0.735 & 8.88E-16 && 0.297 & 1.75E-3 \\
NGC 4526 Center (GC) & Milky Way Center (GC) & 0.140 & \textbf{0.152} && 0.138 & \textbf{0.163} && 0.276 & 9.94E-05 && 0.144 & \textbf{0.131} \\
NGC 4826 (S) & NGC 5253 (S) & 0.129 & \textbf{0.693} && 0.454 & 1.35E-05 && 0.602 & 8.45E-10 && 0.849 & 7.77E-16 \\
Henize 2-10 (S) & Milky Way Center (GC) & 0.403 & 6.05E-05 && 0.131 & \textbf{0.624} && 0.755 & 1.01E-17 && 0.479 & 7.21E-07\\
Henize 2-10 (S) & NGC 4526 Center (GC) & 0.543 & 6.38E-08 && 0.22 & \textbf{0.118} && 0.934 & 4.14E-27 && 0.543 & 6.38E-08 \\
\enddata
\tablecomments{P-values above the 0.05 significance level are bolded.}
\end{deluxetable}
\clearpage
\bibliography{bibliography}{}
\bibliographystyle{aasjournal}

\section{Spatial and Spectral Resolution\label{sec:res}}
Because we are using data sets from a range of resolutions to compare with the Antennae data, it is important to determine how the different resolutions might affect the results. Although deconvolving the data sets (Equations \eqref{eq:deconv} and \eqref{eq:veldeconv}) helps to mitigate the effects of the chosen beam and spectral windows of the observations, it cannot bring out details that the resolution was not high enough to identify in the first place. In order to test some of the differences within our chosen range of resolutions we calculated the physical properties of the Antennae overlap and Henize 2-10 in both their native and smoothed resolution.

\begin{figure}
\centering
\includegraphics{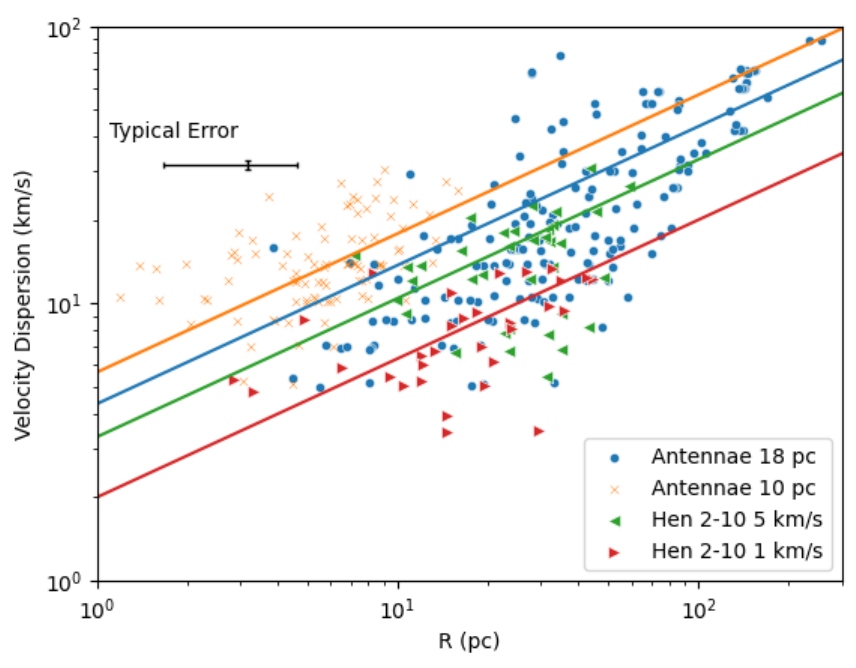}
\caption{Size-linewidth plot showing the effects of different spectral and spatial resolutions. The Antennae overlap is shown at both the 10 and 18 pc spatial resolutions while their spectral resolutions remain at 5 km/s. Contrastingly, Henize 2-10 is kept at 22 pc resolution while spectral resolution is shown at 1 and 5 km/s.}
\label{fig:res}
\end{figure}

Figure \ref{fig:res} plots the effects on the size-linewidth relations, since both the radii and velocity dispersions are affected by different resolutions. \textcolor{black}{For this comparison, we use the same fitting method as Figure \ref{fig:slw1} where the slope is held constant at 0.5 and the intercept is fit using a non-linear least squares method.} While the different resolutions do somewhat affect the derived properties, the differences within the ranges that we have chosen do not significantly change the interpretations. Even though a smaller spatial resolution will result in smaller calculated radii and thus shifting the points farther left on the plot, the measured linewidths are also smaller and the fitted intercepts are only a factor of 2 difference. Similarly, smaller spectral resolution in Henize 2-10 results in both smaller radii and linewidths, and thus their linewidths per size scale are less than a factor of 2 ($\mathrm{\sim}$1.5) from each other.

The trends we have observed between the different environments are on the scale of orders of magnitude. This is much larger than the factors of difference between resolutions in the ranges that we have chosen so the conclusions would remain the same.

\begin{figure}
\centering
\includegraphics[width=1.01\textwidth]{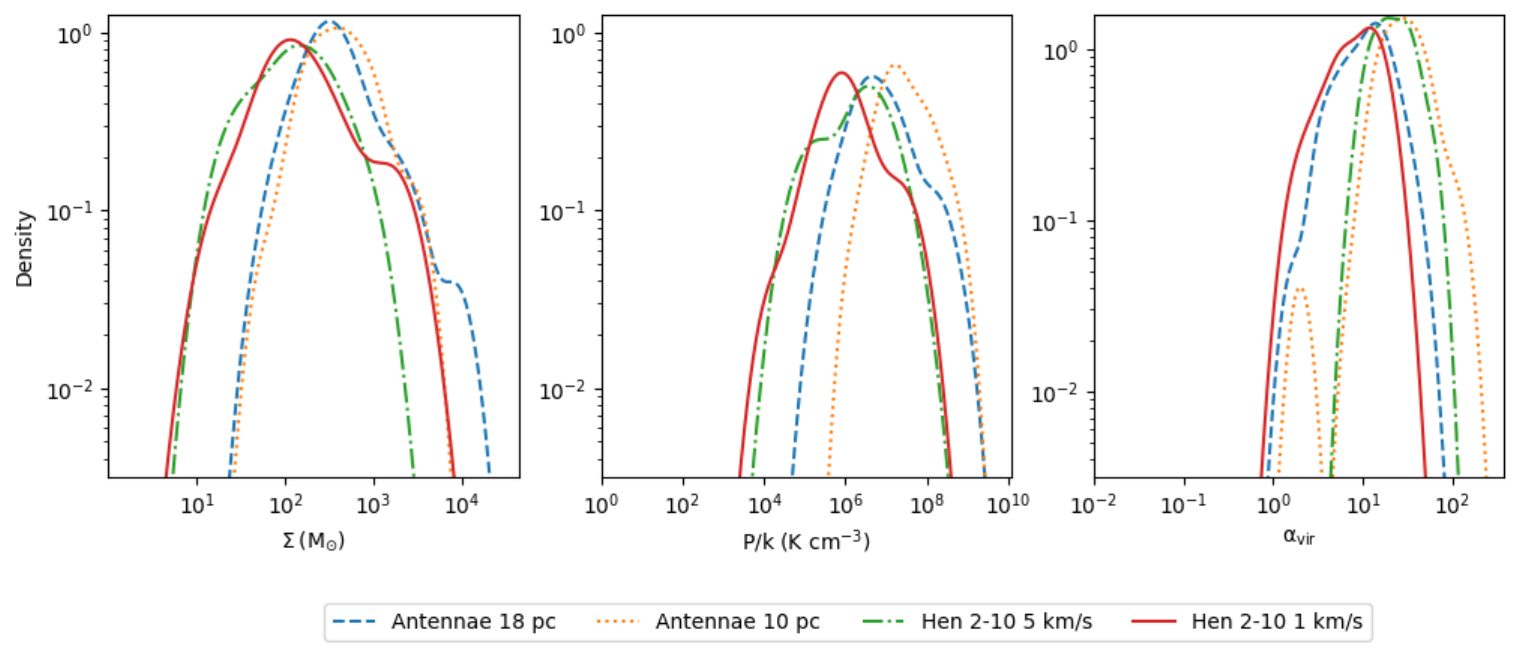}
\caption{{The effects of spatial and spectral resolution on surface density, pressure, and virial parameter (from left to right) distributions for Henize 2-10 and the Antennae overlap.}}
\label{fig:resdist}
\end{figure}

The distributions of other properties such as density, pressure, and virial parameter are similarly affected by resolution effects (Figure \ref{fig:resdist}). There are some differences between the property distributions due to the resolutions, particularly in the pressure distributions, however, the differences are much smaller than the orders of magnitude trends observed in Section \ref{sec:propdist}.

\end{document}